\documentstyle[preprint,aps,eqsecnum,floats,epsf]{revtex}

\tighten

\begin{document}

\def\beq{\begin{equation}}
\def\eeq{\end{equation}}

\def\beqn{\begin{eqnarray}}
\def\eeqn{\end{eqnarray}}
\def\nn{\nonumber\\}

\def\vec#1{{\mbox{\boldmath $\bf #1$}}}

%% scalar products
\def\qe{\e\cdot\vec q}
\def\qk{\vec k \cdot\vec q}
\def\e{\vec\epsilon}

% scalar products with sigma:
\def\Sq{\vec\sigma\cdot\vec q}
\def\Sk{\vec\sigma\cdot\vec k}
\def\Se{\vec\sigma\cdot\vec \e}

\title {The reaction $^2H(\gamma,\pi^0)np$ in the threshold region}

\author
{
M.I. Levchuk$^a$  \thanks{E-mail: levchuk@dragon.bas-net.by},
M. Schumacher$^b$ \thanks{E-mail: schumacher@physik2.uni-goettingen.de}
and 
F. Wissmann$^b$   \thanks{E-mail: fwissma@gwdg.de}
}

\address
{ ~
\\ $^a$B.I. Stepanov Institute of Physics,
   Belarus National Academy of Sciences,
\\ F. Scaryna prospect 70, Minsk 220072, Belarus
}

\address
{ ~
\\ $^b$II Physikalisches Institut der Universit$\ddot a$t G$\ddot 
o$ttingen,
\\ Bunsenstra\ss e 7-9, D-37073 G$\ddot o$ttingen, Germany
}

\maketitle

\begin{abstract}
Neutral pion photoproduction on the deuteron in the inelastic channel 
is studied in the threshold region from 142 to 160 MeV. The 
calculation is based on the use of the diagrammatic approach.
Contributions from the pole diagrams, one-loop  diagrams both with 
$n$-$p$ and $\pi$-$N$ rescattering as well as two-loop diagrams with 
simultaneous inclusion of these rescattering mechanisms are taken 
into account. We have found a reasonable agreement with a very 
simple calculation by J.C.  Bergstrom et al.  [Phys. Rev.  C 57 
(1998) 3203]. It is shown that the `effective' electric dipole 
deuteron amplitude $E_d$ is negative in agreement with a 
prediction of ChPT. We conclude that a deviation of about 
20\% from a ChPT prediction for $E_d$ discovered 
in an experiment on coherent $\pi^0$ photoproduction off the deuteron 
cannot be attributed to inadequate estimates of inelastic channel 
contributions. We suggest that $\pi$-$N$ 
rescattering is responsible for the deviation of the free-nucleon 
$p_1$ threshold amplitude from the value measured in that experiment.

\end{abstract}

\bigskip\noindent
{\it PACS}: 25.20.Lj; 25.45.De

\bigskip\noindent
{\it Keywords}: Pion photoproduction; Low-energy parameters; 
Deuteron

\section{Introduction}
\label{intr}

Nowadays, the threshold $s$- and $p$- amplitudes of $\pi^0$ 
photoproduction off the deuteron are of considerable 
theoretical and experimental interest. This is motivated by two 
reasons.  Firstly, the deuteron is a natural `source' of neutrons.  
If one wants to minimize uncertainties stemming from nuclear forces
when extracting neutron threshold parameters one should use a 
deuteron target. At present, there is a significant disagreement for
values of the threshold electric dipole amplitude $E^{n\pi^0}_{0+}$ 
calculated in chiral perturbation theory (ChPT) and dispersion theory
(DR), respectively, 2.13 \cite{bernard} and 1.19 \cite{hainstein} (in 
units of $10^{-3}/\mu_{\pi^+}$ which are suppressed from here on). 
At the same time the predicted values for $E^{p\pi^0}_{0+}$ agree 
reasonably with each other ($-1.16$ and $-1.22$, respectively) and 
with experimental values $-1.32\pm 0.08$ \cite{bergs96} and $-1.31\pm 
0.08$ \cite{fuchs}.  Resolving the problem mentioned in favour of 
the one or the other theory requires experimental measurements of 
which in the reactions $\gamma d\to\pi^0 d$ and $ed\to e'\pi^0 
d$ seem to be most promising. However,  even in these simplest 
reactions many theoretical problems in the extraction of neutron 
data are anticipated.

On the other hand, the deuteron cannot be considered as a direct sum 
of the proton and neutron (impulse approximation). Photons 
may also interact with potential (virtual) mesons. This 
effect is known as `Meson Exchange Currents' (MEC). It was found to 
be very important in the case of coherent neutral pion 
photoproduction on the deuteron \cite{beane}. Accepting the ChPT 
predictions for free nucleons we could expect in the impulse 
approximation a value of about $+0.5$ ($=\frac 12 
[E^{p\pi^0}_{0+}+E^{n\pi^0}_{0+}]$) for the deuteron electric dipole 
 amplitude $E_d$. In fact, however, the ChPT result is quite 
different being about $-1.8\pm 0.2$ \cite{beane} and the difference 
is due to the meson exchange contributions.

A very recent measurement of $E_d$ performed at SAL \cite{bergstrom}
in coherent $\pi^0$ photoproduction on the deuteron
within 20 MeV of threshold has given a value of $-1.45\pm 0.09$
which is about 20\% lower than the ChPT prediction. 
There is also a significant deviation from the ChPT prediction 
for the free-nucleon $p_1$ amplitude. A comparison of the 
differential cross sections calculated in  a theoretical model 
\cite{benmer} with the data revealed a noticeable disagreement as 
well.  It was claimed in Ref.\ \cite{benmer}  that a possible reason 
for the disagreements might consist in inadequate treatment of the 
inelastic channel in Ref.\ \cite{bergstrom}.  Since the
$^2H(\gamma,\pi^0)np$ reaction cannot be resolved  one has to 
estimate its contribution to the coherent channel making the use of 
theoretical predictions.  Indeed, a model developed in Ref.\ 
\cite{bergstrom} seems to be oversimplified.  For example, the 
authors employed a square-well deuteron wave function and used the 
effective range approximation to describe the final state $n$-$p$ 
interaction.  The deuteron $d$-wave was ignored. Explicit 
calculations of very important diagrams with $\pi$-$N$ rescattering 
were not carried out.  Effectively these latter where included into 
the model by accepting the extracted value $-1.45$ for $E_d$ which, 
as it has been noted, contains contributions from MEC or, in 
other words, from $\pi$-$N$ rescattering. Although a theoretical 
uncertainty of $\pm 25\%$, based on input parameters, was assigned in 
Ref.\ \cite{bergstrom} to the calculated cross sections it would be 
desirable to have more realistic calculations for the inelastic 
channel.

The present paper is aimed at fixing the defects of the model 
\cite{bergstrom}. We are going to study whether these defects may be 
responsible for the deviations mentioned. In our previous paper 
\cite{lps96} on the reaction
\beq
\label{react}
\gamma d\to\pi^0 np
\eeq
we restricted ourselves to the energy region  200 to 400 MeV. Now we 
extend the model to the threshold region.  The extension includes: i) 
a more realistic treatment of the $\pi^0$ photoproduction amplitudes 
at threshold energies; ii) in Ref.\ \cite{lps96}  we considered 
$\pi$-$N$ rescattering mechanism but without 
taking into account the $n$-$p$ rescattering process 
which may follow  $\pi$-$N$ rescattering.
In the present paper a corresponding diagram has been 
taken into account. We would like to emphasize that the study of the 
process (\ref{react}) in the threshold region is hardly of 
theoretical interest in its own. Rather it is needed for 
experimentalists to estimate the relative contribution of the 
inelastic channel to the cross section of coherent pion 
photoproduction on the deuteron. Furthermore, these studies may be 
useful for extracting  the elementary $E^{n\pi^\circ}_{0+}$ amplitude 
from exclusive inelastic data where the recoil neutron is detected.

\section{Kinematics}
\label{kinema}

Let us denote by $k=(\omega,\vec k),~p_d=(\varepsilon_d,-\vec 
k),~q=(\varepsilon_\pi,\vec q),~p_n=(\varepsilon_n,\vec p_n)$ and 
$p_p=(\varepsilon_p,\vec p_p)$ the 4-momenta of the initial photon 
and deuteron and the final pion, neutron and proton, respectively in 
the $\gamma$-$d$ c.m. frame.  A symbol $E_\gamma$ we 
reserve for the lab photon energy $E_\gamma=W_{\gamma d}\omega/m_d$ 
with $W_{\gamma d}=\omega+\varepsilon_d=\omega+\sqrt{\omega^2+m_d^2}$ 
and $m_d$ being the deuteron mass. It is convenient to take as 
independent kinematical variables the photon energy and pion angle, 
$\Theta_\pi$, in the c.m. frame and momentum $\vec p$ of one of the 
nucleons - say, the neutron - in the final $n$-$p$ c.m. frame (see 
Ref.\ \cite{bergstrom}).  Once these variables have been specified 
the pion momentum can be easily found:  
\beqn 
|\vec q|=\frac 1{2W_{\gamma d}} 
\sqrt {[W^2_{\gamma d}-(W_{np}+\mu)^2]
       [W^2_{\gamma d}-(W_{np}-\mu)^2] }, 
\eeqn 
where $W_{np}=2\sqrt{{\vec  p}^2+m^2}$ ($m$ is the nucleon mass) 
and $\mu$ is the $\pi^0$ mass.  Then making a boost with the 
velocity $-\vec q/(W_{\gamma d}-\varepsilon_\pi)$ we 
obtain the neutron momentum $p_n$ and, therefore, totally restore the 
kinematics in the $\gamma$-$d$ c.m. frame. Note that in 
kinematic calculations a distinction is not made between the 
proton and neutron masses. It is, however, taken into account 
in the threshold $\pi^0$ photoproduction amplitude when 
parametrizing the $s$- and $p$-waves multipoles  (see Sect.  
\ref{theory}, Eq.  (\ref{benmer})).

The differential cross section is
\beqn
\label{dcs4}
\frac {d^4\sigma}{d\vec p d\Omega_\pi}=
\frac 1{(2\pi)^5} 
\frac {2m^2\varepsilon_d |\vec q|^3}
{4\omega W_{\gamma d}  W_{np}
(\varepsilon_\pi q \cdot p_p-\varepsilon_p\mu^2)}
\frac 16
\sum _{m_pm_n\lambda m_d} 
| \langle m_pm_n| T| \lambda m_d \rangle |^2,
\eeqn
where $m_p$, $m_n$, $\lambda$, and $m_d$ are spin states of 
the proton, neutron, photon, and deuteron, respectively.

Assuming that near threshold energies the nucleons are not 
detected we should integrate the l.h.s. of Eq.\ (\ref{dcs4}) 
over the momentum $\vec p$ to obtain
\beqn
\label{dcs}
\frac {d\sigma}{d\Omega_\pi}= \int \limits_0^{p^{max}}
\frac {d^4\sigma}{d\vec p d\Omega_\pi} p^2dp d\Omega _{\vec p},
\eeqn
where the maximum value $p^{max}$ is found
from energy-momentum conservation and reads
\beqn
\label{pmax}
p^{max}
=\frac 12
\sqrt {(W_{\gamma d}-\mu)^2-4m^2}. 
\eeqn

\section{Theoretical model for the $\gamma d\to \pi^0 np$ 
process}
\label{theory}

\begin{figure}[htb]
\centerline{\epsfbox[30 430 550 810]{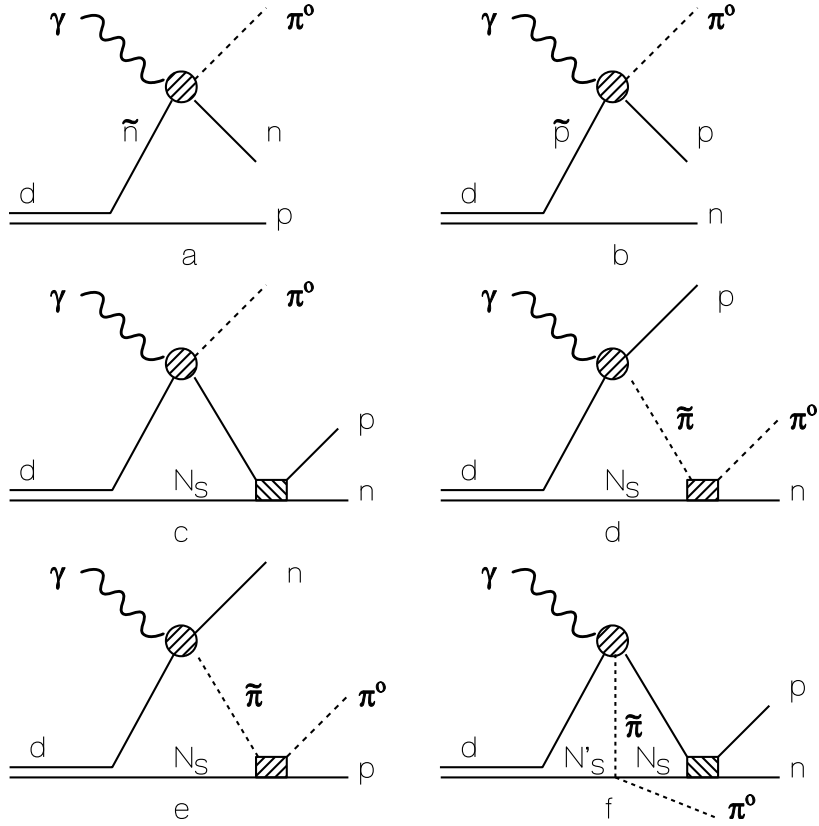}}
\caption{Diagrammatic representation of the scattering 
amplitude.}
\label{fig1}
\end{figure}

The scattering amplitude of the reaction (\ref{react}) is described 
as a sum of contributions from diagrams which are expected to be most 
important in the energy region under consideration. The pole 
diagrams {\it a} and {\it b} in Fig.\ \ref{fig1} and one-loop diagram 
\ref{fig1}{\it c} correspond to the  impulse approximation 
without and with $n$-$p$ interaction in the final state, 
respectively. The necessity to consider the latter mechanism 
follows from the very strong nucleon-nucleon interaction 
at small energies. It might  seem that  diagrams  \ref{fig1}{\it 
d} to {\it f} with $\pi$-$N$ rescattering could be disregarded 
in the threshold region since the $\pi$-$N$ scattering lengths are 
about two orders smaller than those for $n$-$p$ scattering.
In fact, however, there is actually no suppression of the 
diagrams \ref{fig1}{\it d} to {\it f} in comparison with  
diagram \ref{fig1}{\it c}.  Indeed, keeping in mind that 
the threshold electric dipole amplitudes $E_{0+}$ for charged 
channels are about 30 times larger in absolute numbers than those for 
the neutral channels one can expect the corresponding contributions 
from $\pi$-$N$ rescattering to be of the same order as the ones from 
final state $n$-$p$ interaction.  This expectation will be confirmed 
below by numerical calculations. A very big effect of $\pi$-$N$ 
rescattering in the case of coherent $\pi^0$ photoproduction on the 
deuteron in the threshold region  was also found in Refs.\ 
\cite{koch,bosted,benmer}.  Note that our treatment of $\pi$-$N$ 
contribution is similar to the one of `three-body interactions to 
order $q^3$' from Ref.\ \cite{beane}  in the case of coherent   
$\pi^0$ photoproduction off the deuteron. It was found in that paper 
that three-body contributions at order $q^4$ are less important. 
Based on this finding we do not take into account the corresponding 
diagrams for the reaction (\ref{react}) though they of course exist.

We  begin with  the pole diagrams. The matrix element corresponding 
to the diagram in Fig.\  \ref{fig1}{\it a} reads
\beq
\label{pol_n}
\langle m_pm_n| T^{1a}(\vec k,\vec q,\vec p_p) 
| \lambda m_d\rangle =\sum _{m_{\tilde n}}
\Psi ^{m_d}_{m_pm_{\tilde n}}\left({\vec p}_p+\frac {\vec k}2 \right)
\langle m_n| 
T_{\gamma \tilde n\to \pi^0 n}(\vec k_{\pi n},
                               \vec q_{\pi n})
| \lambda m_{\tilde n} \rangle, 
\eeq 
where $\Psi ^{m_d}_{m_pm_{\tilde n}}({\vec 
p}_p+\vec k/2)$ is the deuteron wave function (DWF) and 
$\langle m_n| T_{\gamma n\to \pi^0 n} | \lambda m_{\tilde n}  
\rangle $ is the amplitude of the elementary process $\gamma  n\to 
\pi^0 n$. The amplitude depends on photon ($\vec k_{\pi n}$) and pion 
($\vec q_{\pi n}$) momenta taken in the c.m. frame of the 
$\pi$-$n$ pair. These momenta can be obtained from the 
corresponding momenta in the $\gamma$-$d$ c.m. frame through a boost 
with the velocity $\vec p_p/(W_{\gamma d}-\varepsilon_p)$.

In calculations we used DWF for non-relativistic  versions of 
the Bonn OBE potential \cite{Mach87,Mach89} (the same models 
were used when solving the Lippmann-Schwinger equation for the 
$n$-$p$ scattering amplitude needed for calculations of diagrams 
\ref{fig1}{\it c} and {\it f}, see below). Analytical 
parametrizations of the $s$- and $d$-wave amplitudes were taken from 
Ref.\ \cite{Lev95}.  We would like to note just here that our results 
are practically independent of the choice of the potentials. 

In our previous paper \cite{lps96} we used the well-known 
Blomqvist-Laget parametrization \cite{BlLag77} of 
the pion photoproduction amplitudes. This model is rather simple and, 
nevertheless, fits reasonably to the experimental data. 
Unfortunately, it is obsolete in the sense that it predicts 
the threshold electric dipole amplitudes for the neutral channels in 
accordance with the classical low-energy theorem, 
$E^{p\pi^0}_{0+}=-2.4$ and $E^{n\pi^0}_{0+}=0.4$, but in disagreement 
with the modern values (see Sect. \ref{intr}). To be consistent with 
these latter, we use in the energy region $\omega \leq 150$ MeV a 
parametrization for the neutral pion photoproduction amplitude 
given in Ref.\ \cite{benmer}. The threshold amplitude in the 
$\gamma$-$N$ c.m. frame can be written in the following form 
\cite{bernard}
\beqn
\label{benmer}
T_{\gamma N\to \pi^0N}=-\frac {4\pi W_{\gamma N}}m 
\left[i \Se ~(E_{0+}+\qk~ p_1) + i\Sk ~\qe ~p_2
+\vec q \cdot (\vec k \times \e)~ p_3\right], 
\eeqn
where $W_{\gamma N}$ is the total energy of the $\gamma$-$N$ system, 
and $p_1$, $p_2$ and $p_3$ are the $p$-waves multipoles. 
The $E_{0+}$ amplitude was taken in the form of Eq. (25) of Ref.\ 
\cite{benmer} with parameters from Table I of the same work. The 
parametrization takes into account the energy dependence of the  
$E_{0+}$ multipole as well as the so-called cusp effect.  Just at 
threshold it reproduces the ChPT and DR values (see Sect.  
\ref{intr}). These parametrizations  were used in the present paper 
to study uncertainties due to the choice of a theoretical model for 
pion photoproduction. As to the $p$-wave amplitudes, they are also 
taken for two models and listed in Table\ \ref{tab1}.  As was shown 
in Ref.\ \cite{benmer}, the parametrizations corresponding to ChPT 
and DR reproduce very well the available data on the energy 
dependence of the real part of $E_{0+}$ as well as the total and 
differential cross sections of the reaction $\gamma p\to \pi^0 p$ up 
to $E_\gamma \simeq 180 $ MeV.

\begin{table}[hbt]
\caption 
{
Threshold $p$-waves amplitudes (units are $10^{-3}/\mu^3_{\pi^+}$).
}
\begin{center}
\begin{tabular}{l|cccccc}
\multicolumn{1}{l|}{} &
   \multicolumn{2}{c}{$p_1$} & \multicolumn{2}{c} {$p_2$} &
   \multicolumn{2}{c}{$p_3$} \\
     &$\pi^0p$ &$\pi^0n$ &$\pi^0p$ &$\pi^0n$ &$\pi^0p$ &$\pi^0n$\\
\hline
ChPT & 10.3    & 7.4  & $-11.0$   & $-8.4$    & 11.7  & 11.1 $^a$ \\
DR   & 10.5    & 7.8  & $-11.4$   & $-8.8$    & 10.2  & ~9.5 $^a$ \\
\end{tabular}
\end{center}
$^a$The value was obtained from isospin relations between four
physical channels.
\label{tab1}
\end{table}

One should emphasize that, although we are going to work  in the 
energy region below 160 MeV, we need parametrizations for the 
photoproduction amplitudes at higher energies too. Such 
energies emerge because of two kinds of boosts. The first one occurs 
at the integration over $\vec p$ in Eq. (\ref{dcs}). Below 160 MeV 
the value of $p^{max}$ does not exceed 125 MeV/c so that 
the corresponding boosts lead to only small energy shifts. But at 
integrations over loops when calculating $n$-$p$ final state 
interaction and $\pi$-$N$ rescattering (see below), energy shifts 
can be much bigger. In addition, when considering  $\pi$-$N$ 
rescattering the pion photoproduction amplitudes for the charged  
channels are also needed.  Both these latter and the $\pi^0$ 
photoproduction amplitudes at photon c.m. energies above 150 
MeV were taken from the Blomqvist-Laget model \cite{BlLag77}.

In full analogy with Eq.  (\ref{pol_n}) we write the matrix element
corresponding to  diagram  \ref{fig1}{\it b} in the form
\beq
\label{pol_p}
\langle m_pm_n| T^{1b}(\vec k,\vec q,\vec p_n) 
| \lambda m_d\rangle =\sum _{m_{\tilde p}}
\Psi ^{m_d}_{m_nm_{\tilde p}}\left({\vec p}_n+\frac {\vec k}2 \right)
\langle m_p| 
T_{\gamma \tilde p\to \pi^0 p}(\vec k_{\pi p},
                               \vec q_{\pi p})
| \lambda m_{\tilde p} \rangle. 
\eeq 
The photon ($\vec k_{\pi p}$) and pion  ($\vec q_{\pi p}$) momenta  
in the c.m. frame of the $\pi$-$p$ pair can be obtained by boosting
the corresponding momenta in the  $\gamma$-$d$ c.m. frame  with the 
velocity $\vec p_n/(W_{\gamma d}-\varepsilon_n)$.

The one-loop diagram in Fig.\  \ref{fig1}{\it c} with $n$-$p$ 
rescattering in the final state is expected to be very important in 
the energy region under consideration. The smallness of the relative  
$n$-$p$ momenta leads to strong  $^1S_0$- and $^3S_1$-wave 
interactions which significantly modify the $n$-$p$ plane wave states 
appearing in the pole diagrams \ref{fig1}{\it a} and {\it b}. The 
matrix element corresponding to diagram \ref{fig1}{\it c} is
\beqn
\langle m_pm_n| T^{1c}(\vec k,\vec q,\vec p_n) 
| \lambda m_d\rangle =
-~m\int \frac {d^3{\vec p}_s}{(2\pi)^3}
\frac 1{p^2_{in}-p^2_{out}-i0}
\nn
\times
\sum _{m_{\tilde p}m_{\tilde n}}
\left[
\langle{\vec p}_{out},m_pm_n| T_{np}| -{\vec p}_{in},
 m_{\tilde p}m_{\tilde n}\rangle\langle
m_{\tilde p}m_{\tilde n}| T^{1a}(\vec k,\vec q,\vec p_s) 
| \lambda m_d\rangle
\right.
\nn
\left.
+ 
~\langle{\vec p}_{out},m_pm_n| T_{np}| {\vec p}_{in},
m_{\tilde p}m_{\tilde n}\rangle\langle
m_{\tilde p}m_{\tilde n}| T^{1b}(\vec k,\vec q,\vec p_s) 
| \lambda m_d\rangle
\right],
\label{np-resc}
\eeqn
where ${\vec p}_{out}=({\vec p}_p-{\vec p}_n)/2$ and ${\vec p}_{in}={\vec 
p}_s+{\vec q}/2$ are the relative momenta  of the $n$-$p$ pair after 
and before scattering, respectively, and 
$\langle{\vec p}_{out},m_pm_n| T_{np} | {\vec p}_{in}, m_{\tilde
p}m_{\tilde n}\rangle$  being the half 
off-shell $n$-$p$ scattering amplitude.  We will not discuss here 
details of computations of the amplitude (\ref{np-resc}) because they 
are given in Ref.  \cite{lps96}. Note only that all partial waves 
with the total angular momentum $J\le 1$ were retained in the 
$n$-$p$ scattering amplitude. We cut off the integration in Eq. 
(\ref{np-resc}) at ${|\vec p_s|}^{max}=500$ MeV/c since it is 
difficult to control the pion photoproduction amplitudes at higher 
momenta.  Such a cutting off has no impact on our final results 
because the integral is mainly saturated at smaller momenta.  For 
example, a variation of ${|\vec p_s|}^{max}$ from 400 to 500 MeV 
changes the calculated differential cross sections by less than 2\%.  
One further remark is concerned with the choice of particle energies 
when integrating over the momenta $\vec p_s$.  In accordance with a 
prescription of Ref.\ \cite{laget81} we will suppose  a spectator 
particle (it is denoted in Fig.\ \ref{fig1}{\it c} by a subscript 
$s$) to be on its mass shell.

The amplitude for the diagram in Fig.\  \ref{fig1}{\it d} reads
\beqn
\langle
m_pm_n| T^{1d}(\vec k,\vec q,\vec p_n)
| \lambda m_d\rangle = \hspace {9cm}
\nn
-\int \frac {d^3{\vec p^\ast_s}}{(2\pi)^3} \sum _{m_1m_2}
\Psi ^{m_d}_{m_1m_2}\left( {\vec p}_s+\frac {\vec k}2 \right)  
\left\{\frac  {\varepsilon^\ast_n+\varepsilon^\ast_s}{2W_{\pi n}}
\frac 1{{p^\ast_s}^2-{p^\ast_{\pi n}}^2-i0}
\right.
\nn  
\times 
\Big[ 
  \langle m_n| 
T_{\tilde \pi^0 n\to \pi^0  n}({\vec p}^\ast_{\pi n},{\vec p}^\ast_s) 
| m_1\rangle
  \langle m_p| 
T_{\gamma p\to \tilde \pi^0 p}({\vec k}_{\tilde \pi p},
                        {\vec q}_{\tilde \pi p}) 
| m_2\rangle 
\nn 
- \langle m_n| 
T_{\tilde \pi^-  p\to \pi^0 n}({\vec p}^\ast_{\pi n},{\vec p}^\ast_s) 
| m_1\rangle
  \langle m_p| 
T_{\gamma n\to \tilde \pi^- p}({\vec k}_{\tilde \pi p},
                        {\vec q}_{\tilde \pi p}) 
| m_2\rangle 
\Big] \Bigg\}.   
\label{pinresc1}
\eeqn
In Eq. (\ref{pinresc1}) the asterisk denotes kinematical variables in 
the $\pi^0$-$n$ c.m. frame. 
$W_{\pi n}=\varepsilon^\ast_\pi+\varepsilon^\ast_n=
\sqrt{{p^\ast_{\pi n}}^2+\mu^2}+\sqrt{{p^\ast_{\pi n}}^2+m^2}$
is the total energy in this frame and
\beqn 
{p^\ast_{\pi n}}^2=\frac 1{4W^2_{\pi n}} 
       [W^2_{\pi n}-(m+\mu)^2]
       [W^2_{\pi n}-(m-\mu)^2] > 0.
\label{p_aster}
\eeqn 
A boost with the velocity $-\vec  
p_p/(W_{\gamma d}-\varepsilon_p)$ is needed to transform 
the momentum  $\vec p^\ast_s$ to $\vec p_s$ in the 
$\gamma$-$d$ c.m. frame. As in the case of Eq. (\ref{np-resc}) we 
cut off the integration  in Eq. (\ref{pinresc1}) at ${|\vec 
p^\ast_s|}^{max}=500$ MeV/c without any noticeable changes in the
calculated differential cross sections.  Spectator particles in 
diagrams \ref{fig1}{\it d} and {\it e} are again supposed to be on 
their mass shells.

The $\pi$-$N$ scattering amplitude  has the following form
\beqn
\label{pinampl}
\langle m'| T^I_{\pi N\to\pi' N'}({\vec q}', {\vec q})| 
m\rangle &=&
(2\pi)^3~ \sqrt { 
\frac {4\varepsilon^\ast_\pi {\varepsilon_\pi'}^\ast
 \varepsilon^\ast_N {\varepsilon_N'}^\ast}{m^2} }
\nn  &&
\times
\sum _{JLm_J} 
 C^{Jm_J}_{ \frac 12 m' L m'_L } Y^{m'_L}_L( {\hat {\vec q}'})
~C^{Jm_J}_{ \frac 12 m  L m_L  } {Y^{m_L }_L}^\ast( {\hat {\vec q}})
~T^L_{IJ}(q',q),
\eeqn
where ${\vec q}'$ and $\vec q$ are the final and initial momenta of
(generally, off-shell) particles in the $\pi$-$N$ c.m. frame, 
$Y^{m_L}_{L}({\hat {\vec q}})$ are the spherical harmonics,
$C^{JM}_{J_1M_1J_2M_2}$ are the Clebsch-Gordan coefficients.
Partial wave amplitudes $T^L_{IJ}(q',q)$ for all $s$- and $p$-waves 
channels were obtained by solving the Lippmann-Schwinger 
equation for a separable energy-dependent $\pi$-$N$ potential
built in Ref.\ \cite{nozawa}. Only for the $p_{33}$-wave we 
preferred using a model from Ref.\ \cite {BlLag77} with explicit 
inclusion of a $s$-channel pole diagram with an intermediate $\Delta$ 
isobar. Contributions from the $d$-waves were found to be 
negligible.  Two $\pi$-$N$ scattering amplitudes in isospin space are 
numbered by a symbol $I=\frac 12$ or $\frac 32$.  Needed in Eq.  
(\ref{pinresc1}) physical amplitudes are linear combinations of 
these isospin amplitudes:                                  
\beqn
T_{\pi^0 p\to \pi^0  p}& =\quad  T_{\pi^0 n\to \pi^0  n}=
\frac 13 \left( T^{\frac 12}+2T^{\frac 32}\right), 
\nn
T_{\pi^+  n\to \pi^0 p}& =- T_{\pi^-  p\to \pi^0 n}=
\frac {\sqrt{2}}3\left(T^{\frac 12}-T^{\frac 32}\right).
\label{ispinpin}
\eeqn

In full analogy with Eq. (\ref{pinresc1}) one has for diagram   
\ref{fig1}{\it e}
\beqn
\langle
m_pm_n| T^{1e}(\vec k,\vec q,\vec p_p)
| \lambda m_d\rangle = \hspace {9cm}
\nn
-\int \frac {d^3{\vec p^\ast_s}}{(2\pi)^3} \sum _{m_1m_2}
\Psi ^{m_d}_{m_1m_2}\left({\vec p}_s+\frac {\vec k}2\right) 
\left\{\frac  {\varepsilon^\ast_p+\varepsilon^\ast_s}{2W_{\pi p}}
\frac 1{{p^\ast_s}^2-{p^\ast_{\pi p}}^2-i0}
\right.
\nn
\times 
\Big[ 
  \langle m_p| 
T_{\tilde \pi^0 p\to \pi^0  p}({\vec p}^\ast_{\pi p},{\vec p}^\ast_s) 
| m_1\rangle
  \langle m_n| 
T_{\gamma n\to \tilde \pi^0 n}({\vec k}_{\tilde \pi n},
                        {\vec q}_{\tilde \pi n}) 
| m_2\rangle 
\nn
- \langle m_p| 
T_{\tilde \pi^+  n\to \pi^0 p}({\vec p}^\ast_{\pi p},{\vec p}^\ast_s) 
| m_1\rangle
  \langle m_n| 
T_{\gamma p\to \tilde \pi^+ n}({\vec k}_{\tilde \pi n},
                        {\vec q}_{\tilde \pi n}) 
| m_2\rangle 
\Big] \Bigg\},   
\label{pinresc2}
\eeqn
where the meaning of all variables is easily understood from the 
corresponding ones in Eq.  (\ref{pinresc1}).

Finally, let us consider  diagram \ref{fig1}{\it f} which includes 
simultaneously both $\pi$-$N$ and $n$-$p$ rescattering. One has for 
the corresponding matrix element

\beqn
\langle m_pm_n| T^{1f}(\vec k,\vec q,\vec p_n)
| \lambda m_d\rangle =
-~m\int \frac {d^3{\vec p}_s}{(2\pi)^3}
\frac 1{p^2_{in}-p^2_{out}-i0}
\nn
\times
\sum _{m_{\tilde p}m_{\tilde n}}
\left[
\langle{\vec p}_{out},m_pm_n| T_{np} | -{\vec p}_{in}, m_{\tilde
p}m_{\tilde n}\rangle\langle
m_{\tilde p}m_{\tilde n}| T^{1d}(\vec k,\vec q,\vec p_s)
| \lambda m_d\rangle
\right.
\nn
\left.
+ 
~\langle{\vec p}_{out},m_pm_n| T_{np} | {\vec p}_{in}, m_{\tilde
p}m_{\tilde n}\rangle\langle
m_{\tilde p}m_{\tilde n}| T^{1e}(\vec k,\vec q,\vec p_s)
| \lambda m_d\rangle
\right].
\label{np-pin-resc}
\eeqn
The amplitude  (\ref{np-pin-resc}) includes as subblocks the 
amplitudes (\ref{pinresc1}) and (\ref{pinresc2}) but with off-shell 
outgoing nucleons so that ${p^\ast_{\pi N}}^2$ in Eq. (\ref{p_aster}) 
may be both positive and negative. One point which has to be 
emphasized is how to specify the energies in the loops. It is 
reasonable to suppose that the nucleon N$'_{\rm S}$ in the left loop 
is on its mass shell since such a choice precisely corresponds to 
that used when evaluating diagrams \ref{fig1}{\it d} and {\it e}.  As 
to the right loop, a prescription for the nucleon energies used in 
the case of coherent $\pi^0$ photoproduction on the deuteron 
consisted in putting the recoil nucleon on its mass shell in the pion 
photoproduction vertex \cite{bosted}.  This prescription (which is 
referred to from here on as Case 1) was motivated by the location of 
the leading singularities of elementary amplitudes (the $\Delta$ pole 
singularity) and internal particle propagators. It was found in Ref.\ 
\cite{bosted} that putting the recoil nucleon on its mass shell at 
the $\pi$-$N$ vertex led to a one and a half times larger 
rescattering amplitude. We think, however, that the prescription 
above can hardly be considered as a hard-and-fast rule.  Even a 
glance on Fig.\ \ref{fig1}{\it f} enables one to see that there are 
no reasons why such a choice of nucleon energies should be 
preferred.  In fact, we have checked other realistic possibilities to 
specify the nucleon energies in the right loop. One of them was to
suppose that each of the nucleons carried equal parts of the total 
energy of the final $n$-$p$ pair, namely 
$(\varepsilon_n+\varepsilon_p)/2$ (Case 2). One more choice rejected 
in Ref.  \cite{bosted} was to put the recoil nucleon on its mass 
shell at the $\pi$-$N$ vertex (Case 3).

Summations over polarizations of the particles in Eqs. 
(\ref{pol_n})-(\ref{np-pin-resc}) as well as the 
three-dimensional integrations in Eqs. 
(\ref{np-resc}), (\ref{pinresc1}) and (\ref{pinresc2}) 
and six-dimensional one in Eq. (\ref{np-pin-resc}) have been 
carried out numerically making use of the methods described in  
Refs.\ \cite{LLP94,Lev95}. Note, since Eq.\ (\ref{dcs}) involves  a 
three dimensional integration we were forced to carry out 
integrations with dimensions up to and including nine. Additional 
one-dimensional integrations were needed to solve the 
Lippmann-Schwinger equations for the $N$-$N$ and $\pi$-$N$ scattering 
amplitudes. The computing work was very hard but we were in a 
position with reasonable computer resources to evaluate the integrals 
with a precision providing an accuracy of calculated cross sections 
better than 10\%.

\section{Results and Discussion}

Based on the approach described in the previous section, we 
calculated the differential cross section (\ref{dcs}) and total 
cross section of the reaction (\ref{react}) in the energy region from 
the threshold ($E^{thr}_\gamma=142.2$ MeV) to 160 MeV.

\begin{figure}[htb]
\centerline{
\leavevmode\epsfxsize=0.40\textwidth \epsfbox{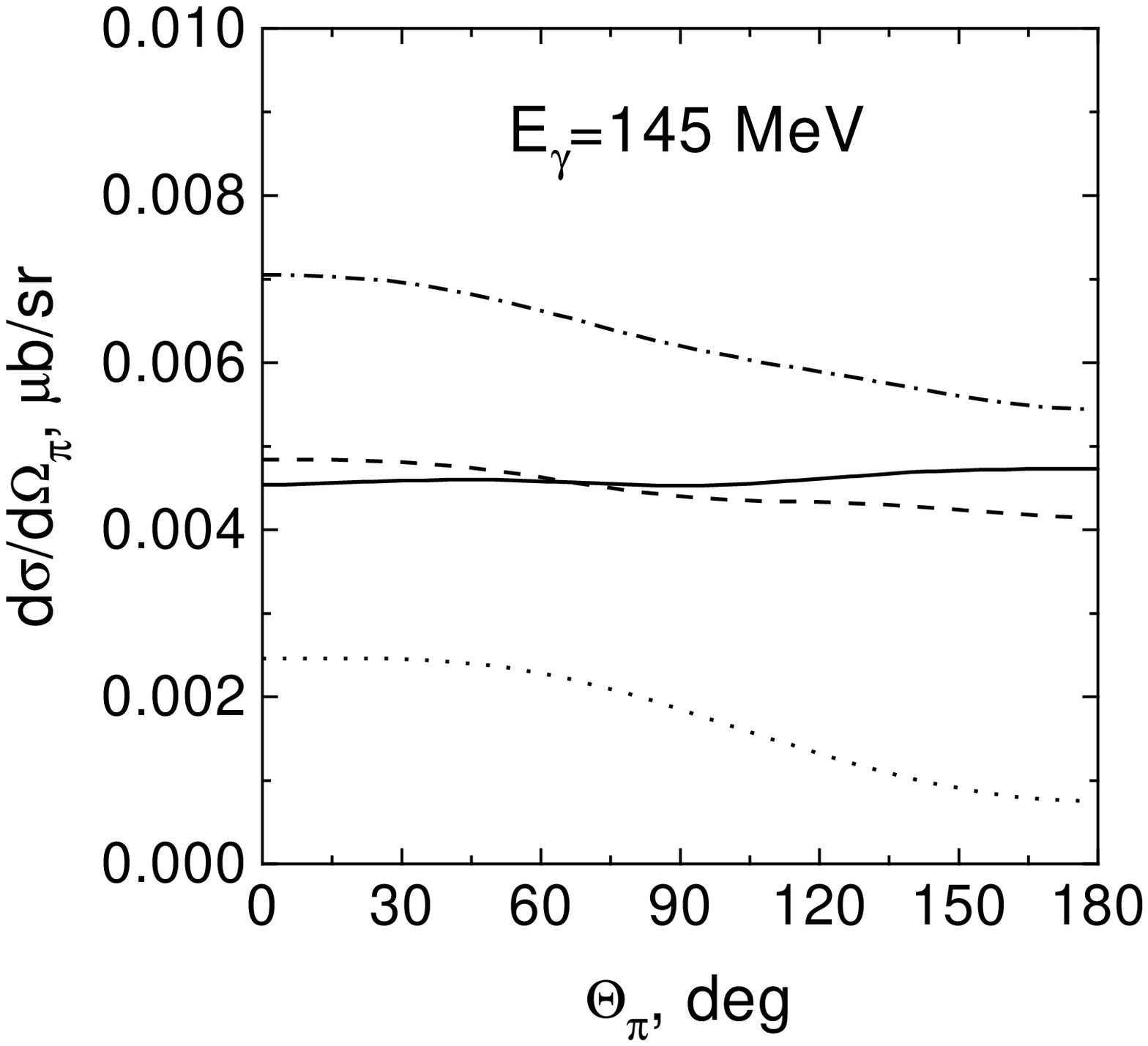}
\leavevmode\epsfxsize=0.40\textwidth \epsfbox{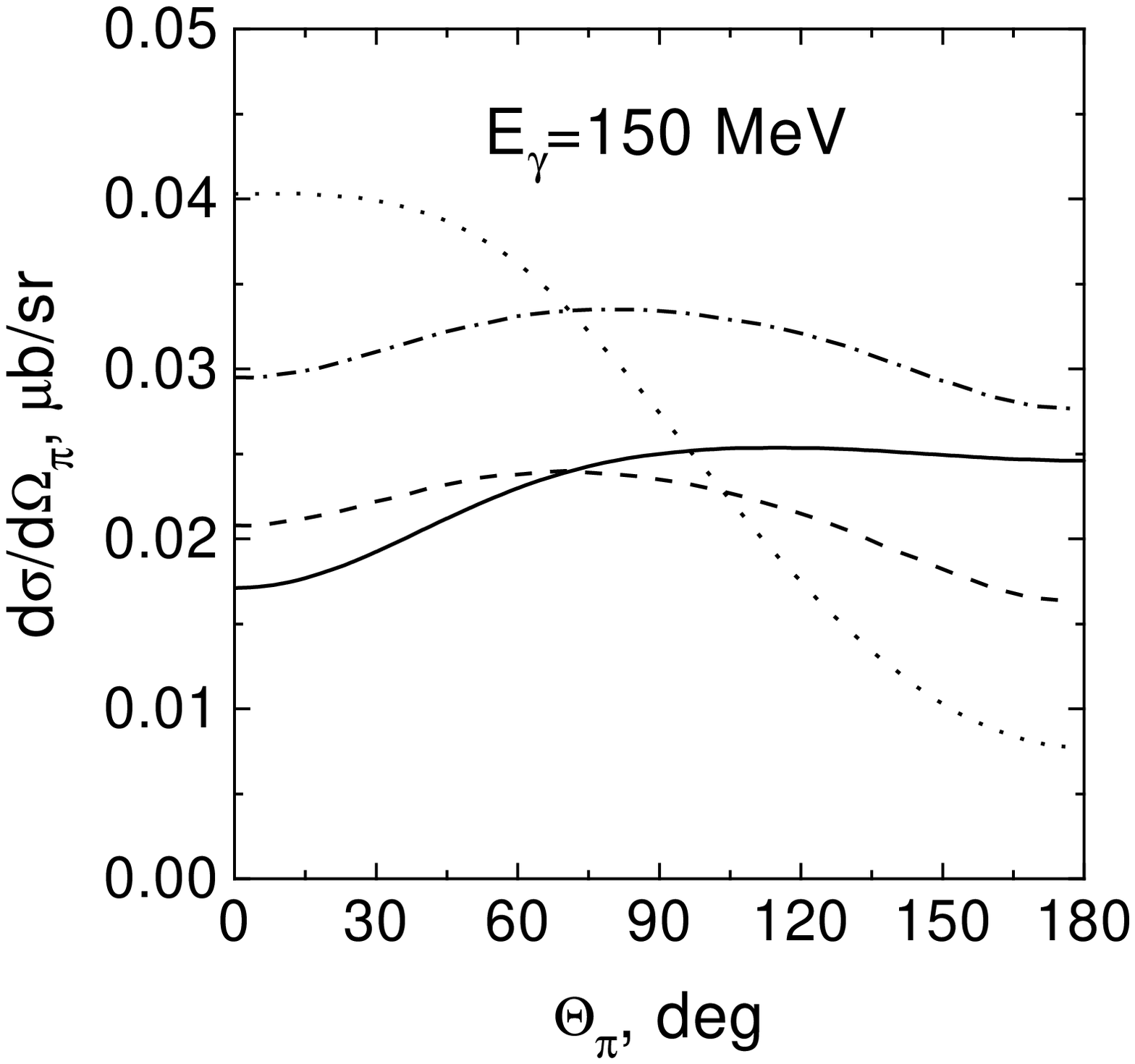} }
\centerline{
\leavevmode\epsfxsize=0.40\textwidth \epsfbox{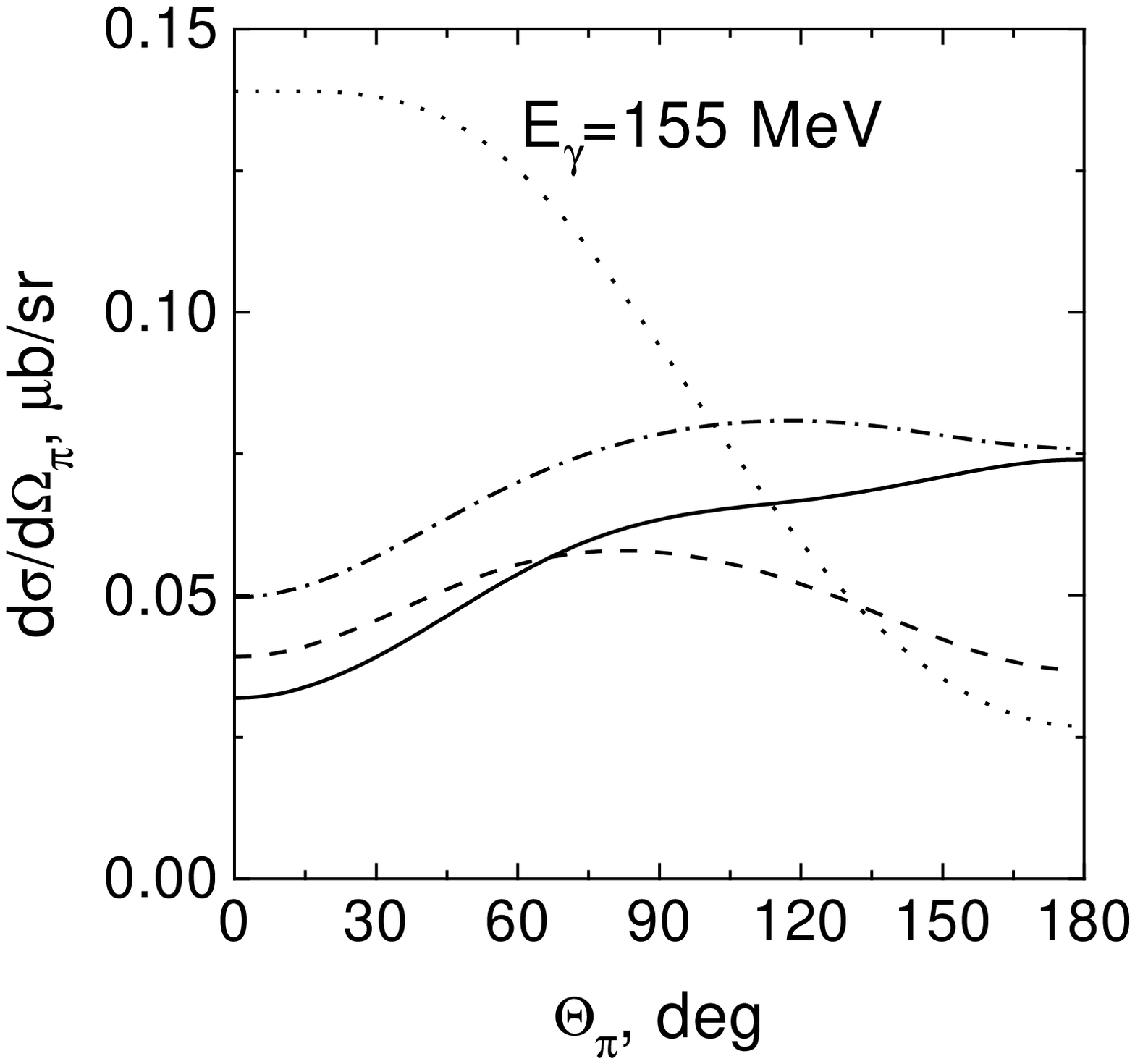}
\leavevmode\epsfxsize=0.40\textwidth \epsfbox{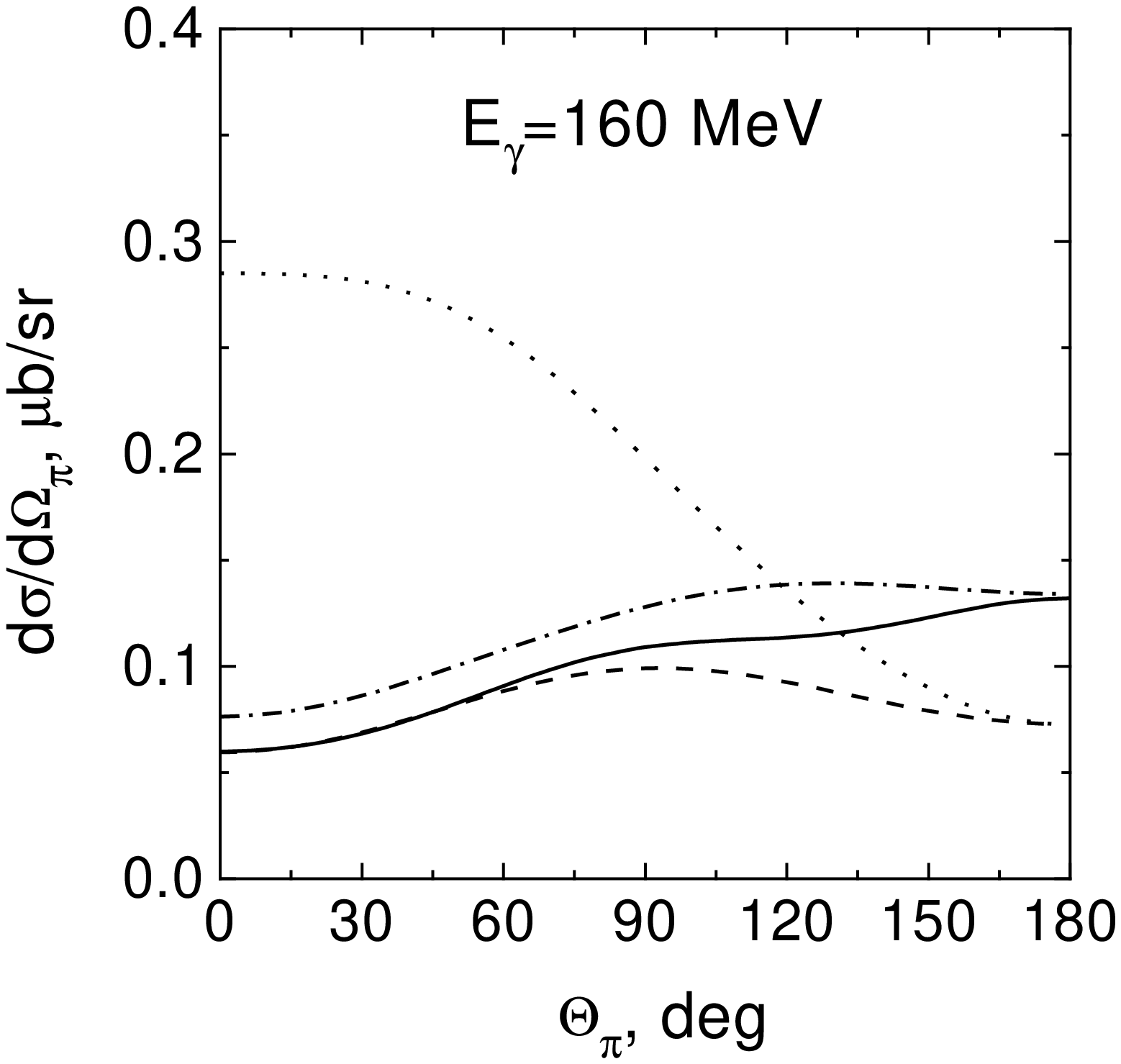} }
\caption{
Shown are contributions of different diagrams to
the differential cross section (\ref{dcs}) in the 
c.m. frame.  Contribution of the pole diagrams \ref{fig1}{\it a} and 
{\it b} is given in dotted lines.  Successive addition of the 
diagrams with $n$-$p$ rescattering (\ref{fig1}{\it c}), $\pi$-$N$ 
rescattering (\ref{fig1}{\it d} and {\it e}) and diagram 
\ref{fig1}{\it f} gives the dashed, dash-dotted and solid lines, 
respectively. The curves correspond to the ChPT set for the 
multipoles in Eq.  (\ref{benmer}) and Case 1 for nucleon energies in 
the right loop of diagram \ref{fig1}{\it f}.
}
\label{fig2}
\end{figure}

In Fig.\ \ref{fig2} we present our predictions for the differential 
cross section with the ChPT set for the multipoles in Eq. 
(\ref{benmer}) and with the choice Case 1 for nucleon energies in 
the right loop of diagram \ref{fig1}{\it f}.  One can see that 
the contributions of all the diagrams in Fig.\ \ref{fig1} are of the 
same order so that none of the diagrams can be ignored in the 
calculation.  The two pole diagrams give a forward-peaked angular 
distribution. This reflects the fact that the ChPT values of the 
isospin averaged threshold electric dipole amplitude 
$E^{(+)}_{0+}=\frac 12 [E^{p\pi^0}_{0+}+E^{n\pi^0}_{0+}]=0.5$ and 
the amplitude $p^{(+)}_1=\frac 12 [p^{p\pi^0}_1+p^{n\pi^0}_1]=8.9$ 
are positive.  Indeed, it is seen from Eq.  (\ref{benmer}) that the 
amplitudes $E^{(+)}_{0+}$ and $p^{(+)}_1$ are responsible for the 
forward-backward asymmetry in the differential cross section.  An 
effect of  $n$-$p$ final state interaction (diagram \ref{fig1}{\it 
c}) is strongly dependent on the photon energy. Just above threshold 
(e.g., at $E_\gamma$=145 MeV) it leads to increasing the differential 
cross sections over the full angular region and this increase is  
attributed to the strong attractive $n$-$p$ interaction in the 
singlet $^1S_0$-state.  At higher photon energies the repulsive 
triplet $^3S_1$-interaction grows in importance.  At $E\geq 150$ MeV 
the total effect of diagram \ref{fig1}{\it c} consists in a 
noticeable lowering of the cross sections except in the 
backward angle region. 

\begin{figure}[htb]
\centerline{
\leavevmode\epsfxsize=0.40\textwidth \epsfbox{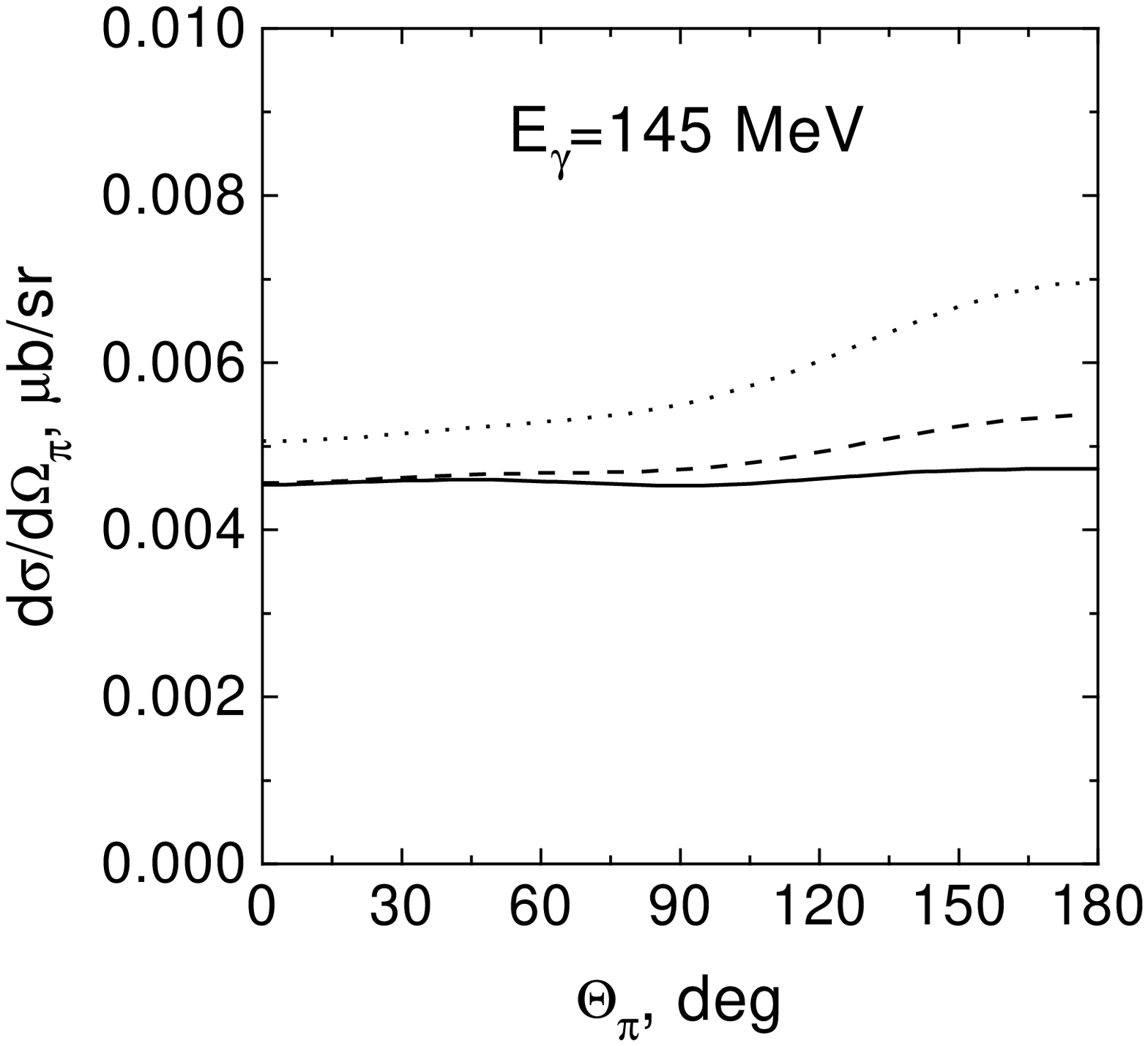}
\leavevmode\epsfxsize=0.40\textwidth \epsfbox{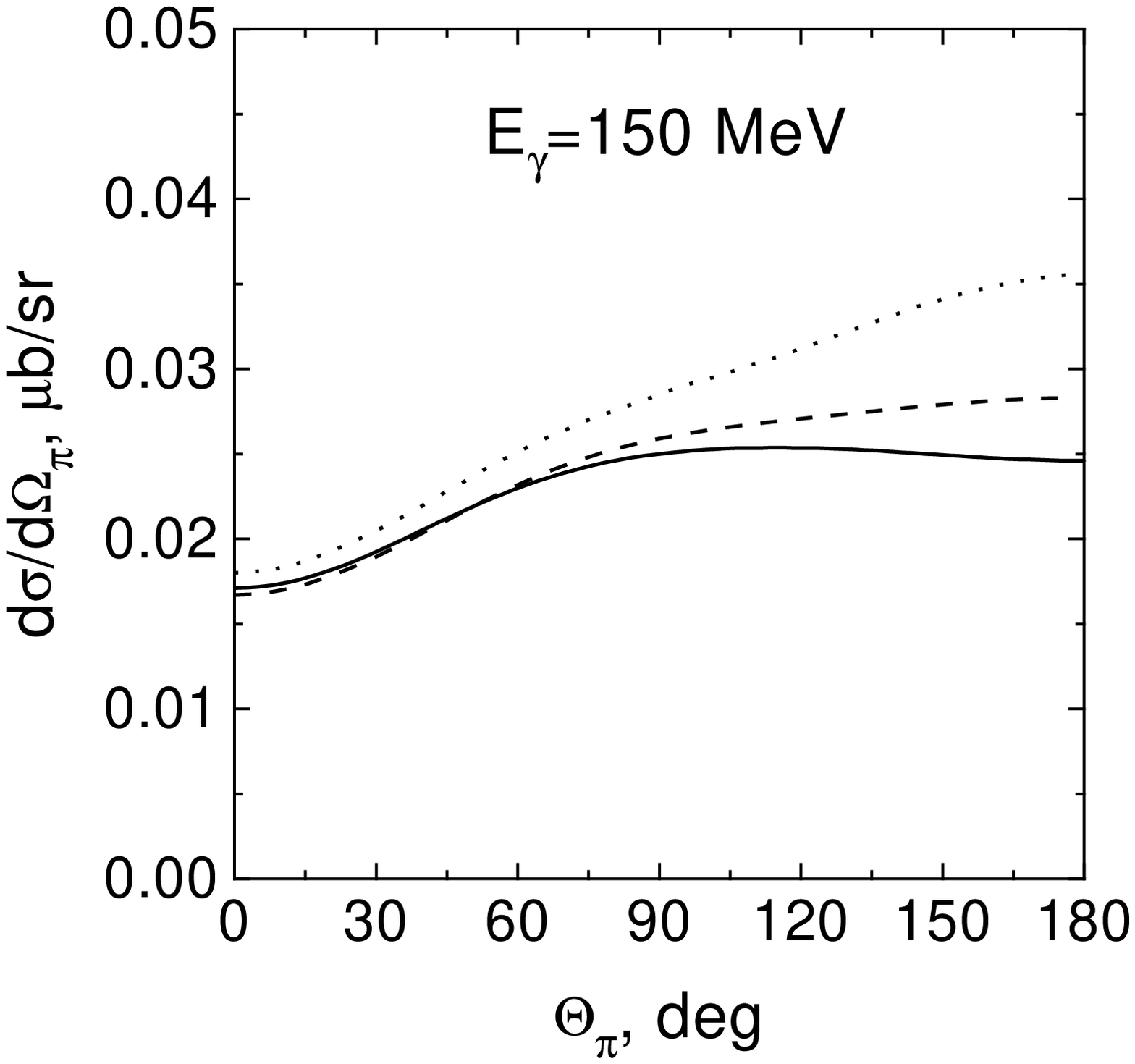} }
\centerline{
\leavevmode\epsfxsize=0.40\textwidth \epsfbox{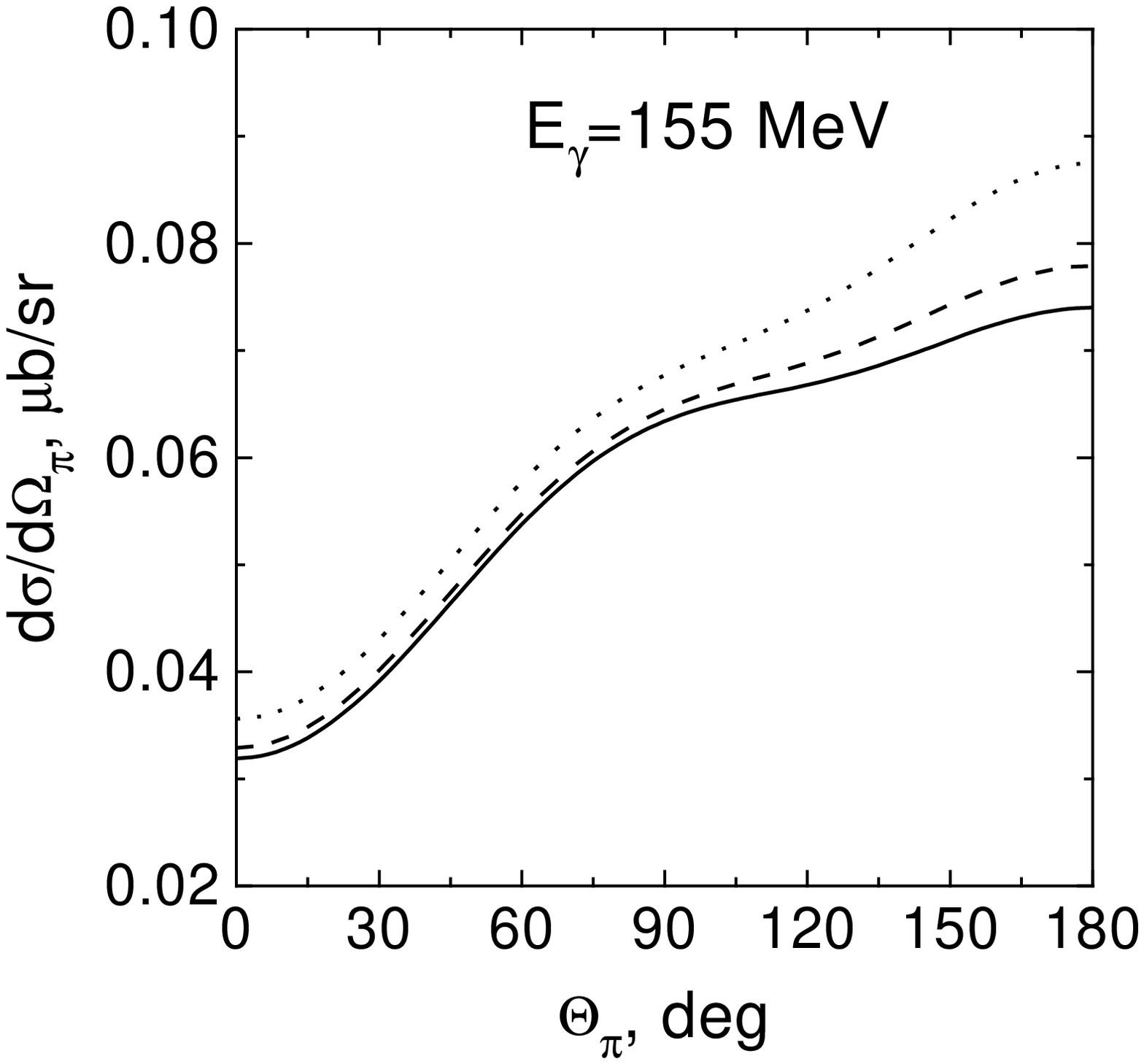}
\leavevmode\epsfxsize=0.40\textwidth \epsfbox{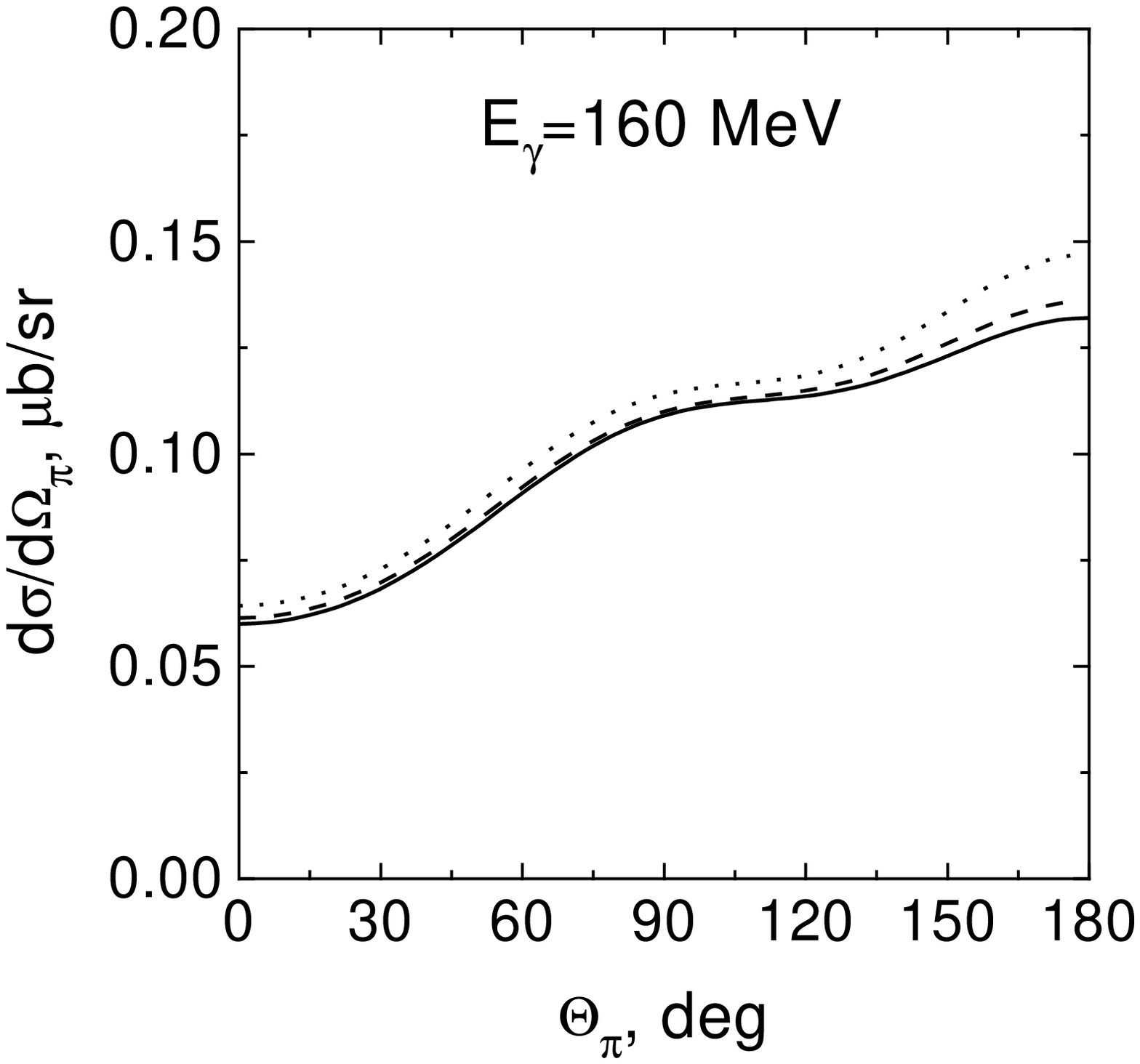} }
\caption
{
Shown is the sensitivity of the 
differential cross section (\ref{dcs}) in the c.m. frame to  
different prescriptions for nucleon energies in the right loop of 
diagram \ref{fig1}{\it f}: Case 1 (solid), Case 2 (dashed), Case 3 
(dotted). The curves  correspond to the ChPT set for the multipoles 
in Eq. (\ref{benmer}).
}
\label{fig3_1}
\end{figure}

As to diagrams \ref{fig1}{\it d} and {\it e} with  $\pi$-$N$ 
rescattering in the final state, they increase the differential cross 
section at all angles. At $E\geq 155$ MeV these diagrams contribute 
mainly at backward angles. Finally, diagram  \ref{fig1}{\it f} gives 
a noticeable reduction of the cross section at forward angles. It is 
seen in Fig.\ \ref{fig2} that after inclusion of diagrams
\ref{fig1}{\it d} to {\it f} the differential cross section becomes 
to be backward peaked at $E\geq 150$ MeV.  This means that due to 
$\pi$-$N$ rescattering the amplitude $E^{(+)}_{0+}$ effectively 
acquires a negative contribution altering its sign. When making such 
a statement we suppose (and with some justice, see below) that 
although rescattering can also modify the $p^{(+)}_1$ amplitude the 
latter remains to be positive.  This effect was first discovered in 
the framework of the ChPT \cite{beane} in the study of coherent 
$\pi^0$ photoproduction on the deuteron. 

With the ChPT set for the multipoles in Eq.  (\ref{benmer}) we 
have calculated and presented in Fig.  \ref{fig3_1} the differential 
cross sections for various choices of nucleon energies in the 
right loop of diagram \ref{fig1}{\it f}.  One can see that the 
results at $\Theta_\pi \le 90^\circ$ are practically independent of 
the choice of the energies. This is not the case for backward angles.  
Case 1 leads to minimum cross sections and Case 3 to maximum ones.  
Results with Case 2 lie between. Simultaneously,  with increasing 
photon energy the deviation is declining. It is worth mentioning that 
analogous observations were made in Ref.\ \cite{benmer} in
consideration of coherent $\pi^0$ photoproduction in the threshold 
region. It was found in that paper that various approximations for 
the nucleon kinetic energy at integrations over loops gave 
noticeable changes in the calculated differential cross sections 
especially near threshold and backward angles.

\begin{figure}[htb]
\centerline{
\leavevmode\epsfxsize=0.40\textwidth \epsfbox{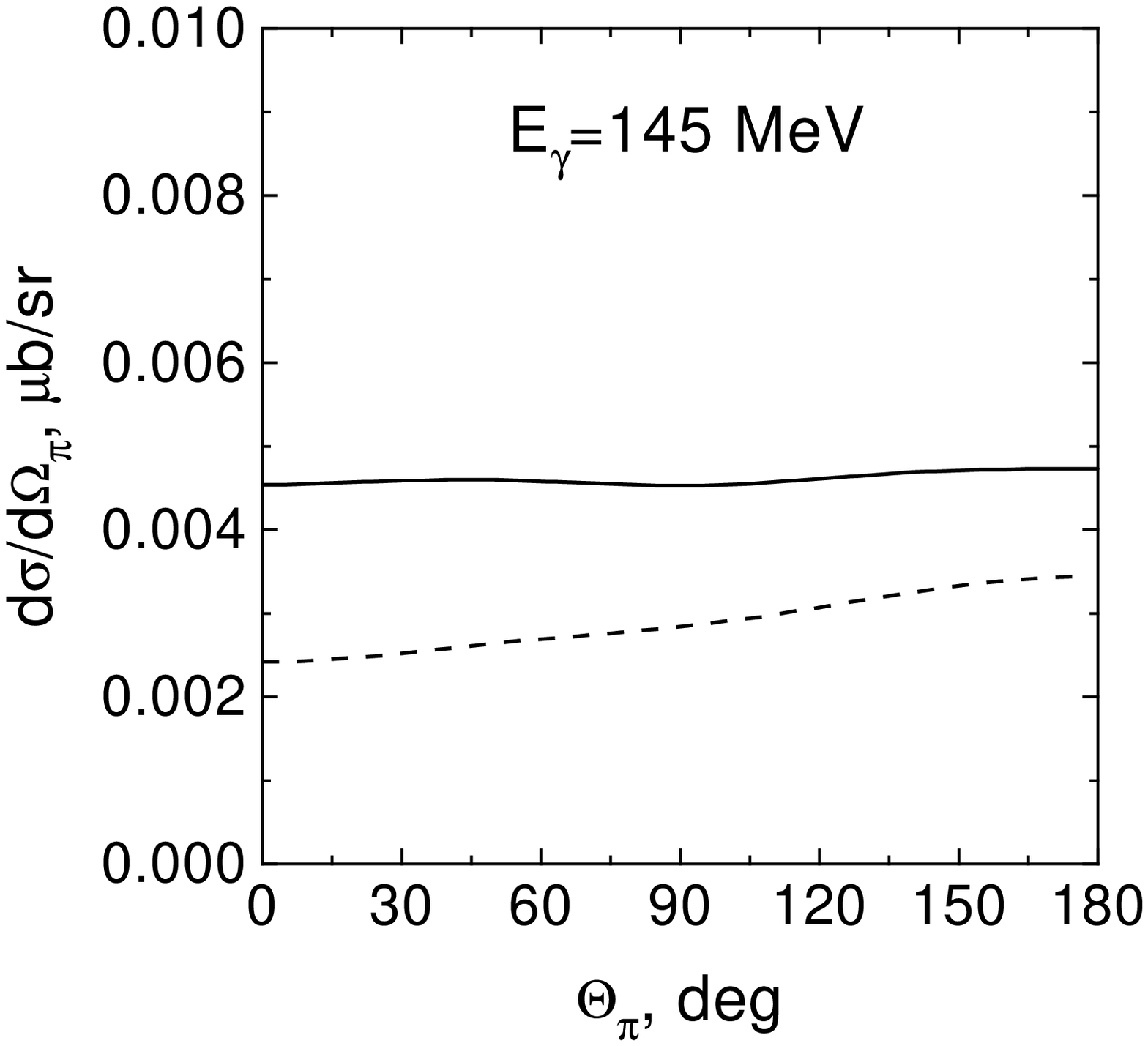}
\leavevmode\epsfxsize=0.40\textwidth \epsfbox{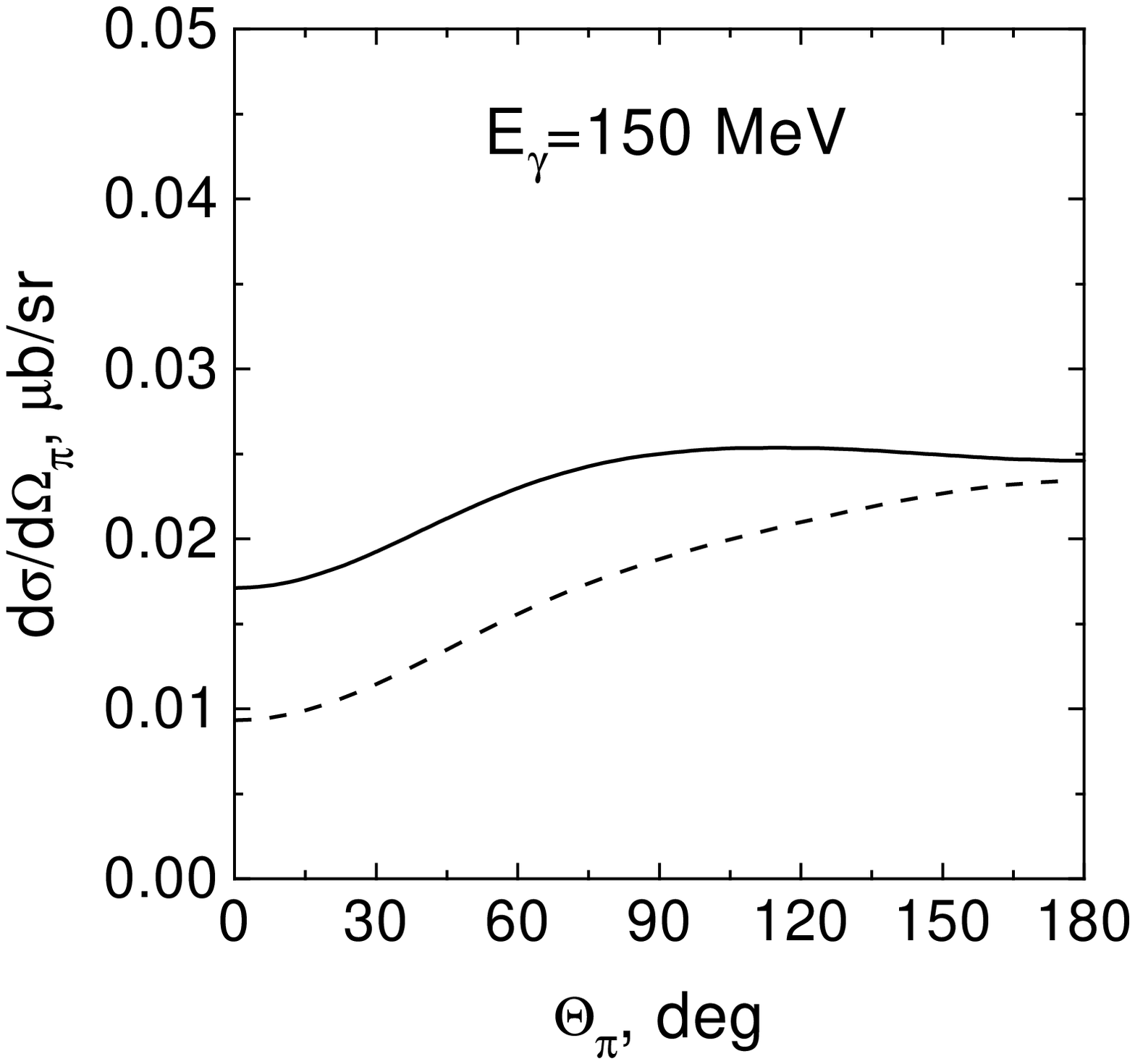} }
\centerline{
\leavevmode\epsfxsize=0.40\textwidth \epsfbox{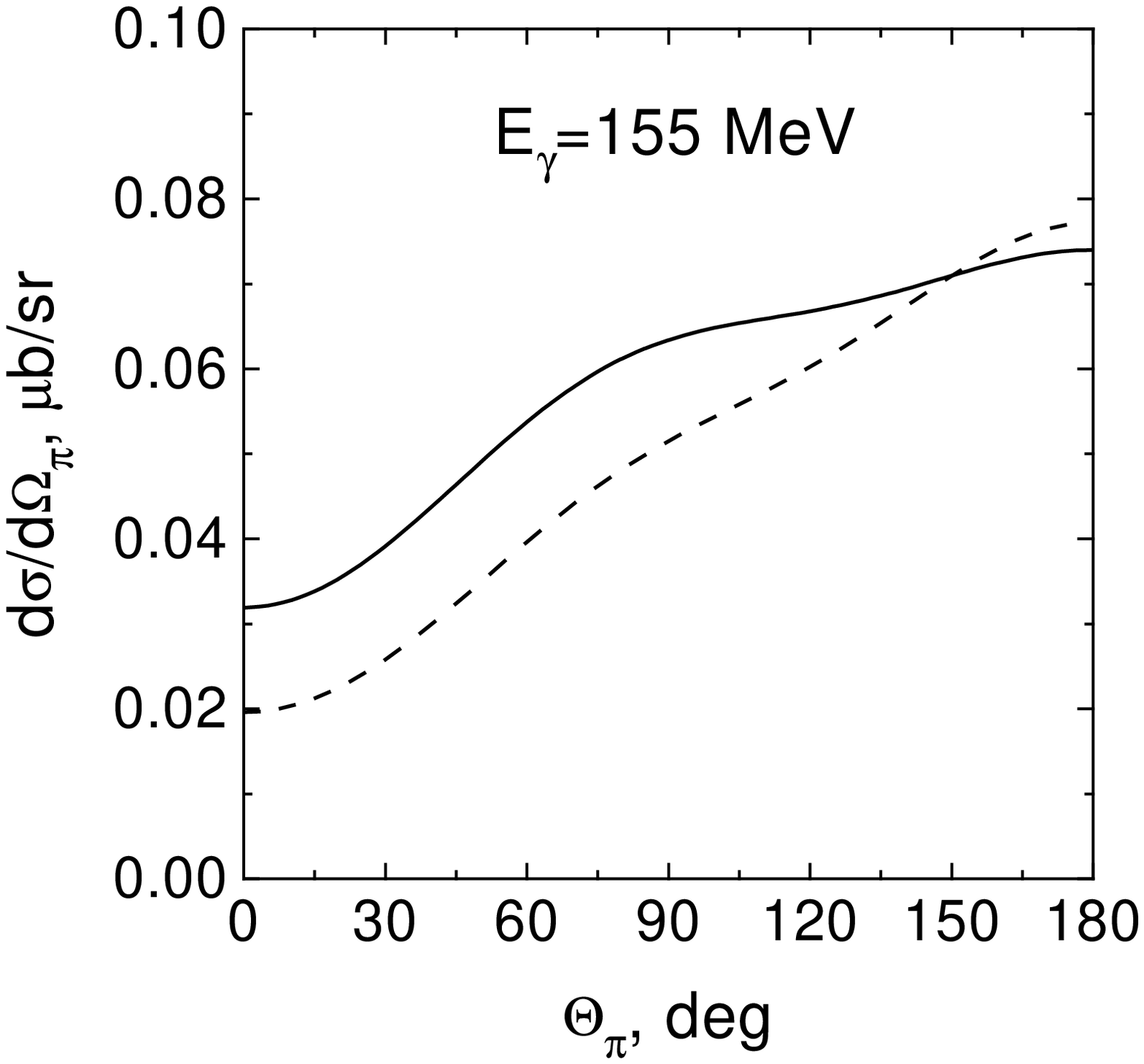}
\leavevmode\epsfxsize=0.40\textwidth \epsfbox{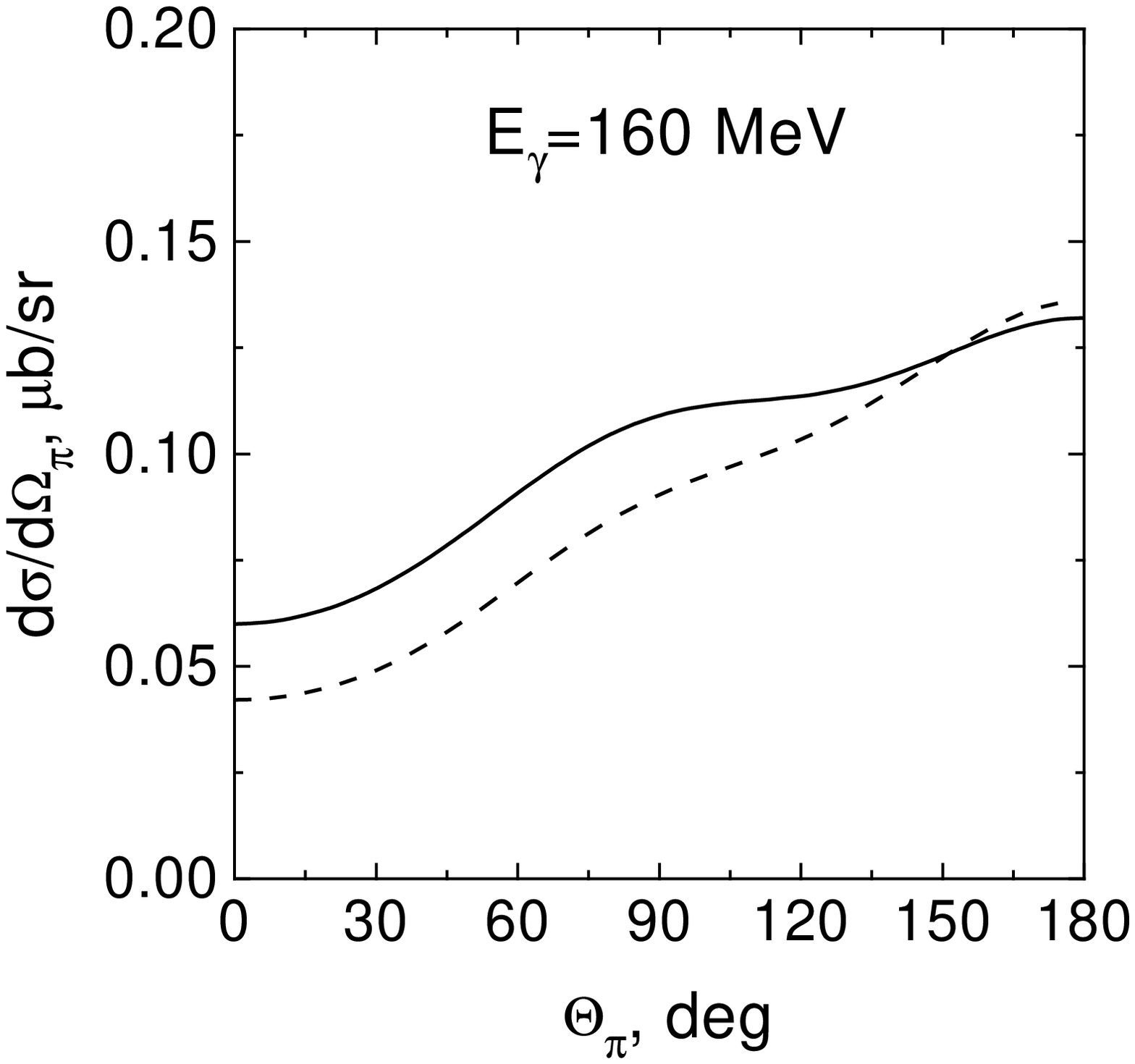} }
\caption
{
Shown is the sensitivity of the 
differential cross section (\ref{dcs}) in the c.m. frame to  
sets of the multipoles in Eq. (\ref{benmer}): DR (dashed), 
ChPT (solid). 
The curves correspond to Case 1 for nucleon energies.
}
\label{fig3}
\end{figure}

In Fig.\ \ref{fig3} we compare our predictions for the angular 
distribution of the differential cross section with the ChPT and 
DR sets for the multipoles. It is seen that the ChPT and DR sets 
produce rather close cross sections at backward angles for 
$E_\gamma \ge 150$ MeV though there exist noticeable 
deviations at forward angles.  In the energy region under 
consideration the DR sets produce only half of the ChPT cross 
sections at forward angles.  This reflects the significant difference 
between the ChPT and DR predictions for the threshold electric 
dipole amplitude $E^{n\pi^0}_{0+}$ (see Sect.  \ref{intr}) and the 
close consistency of the two for the $p^{(+)}_1$ amplitude.   

The previous consideration shows that $\pi$-$N$ rescattering  
(diagrams \ref{fig1}{\it d}-{\it f}) markedly changes the results 
calculated without this effect (diagrams \ref{fig1}{\it a}-{\it c}).  
It is instructive to study the question: what modification of the low 
energy parameters in the model without $\pi$-$N$ 
rescattering is required in order to reproduce the predictions of the 
full model.  We will be interested in the amplitudes  $E^{(+)}_{0+}$ 
and $p^{(+)}_1$ only. Let us suppose that due to $\pi$-$N$ 
rescattering these amplitudes acquire additional contributions 
$\Delta E^{(+)}_{0+}$ and $\Delta p^{(+)}_1$. Numerical estimates 
for them are obtained by switching off diagrams \ref{fig1}{\it 
d}-{\it f} and adjusting these parameters to reproduce the 
differential cross sections at $\Theta_\pi=0^\circ$ and $180^\circ$ 
for the full model. Our results are given in Table\ \ref{tab2}. 
One comment should be made  here.  In a rigorous consideration the 
values $\Delta E^{(+)}_{0+}$ and $\Delta p^{(+)}_1$ must be 
independent of the parametrization for the low energy amplitudes  in 
Eq.\ (\ref{benmer}), i.e. the former should be the same for ChPT 
and DR amplitudes.  Indeed, the difference between ChPT and DR 
manifests itself in the $\pi^0$ production amplitudes but the main 
contribution from $\pi$-$N$ rescattering stems from charged pion 
exchanges. The slight difference for ChPT and DR  results in Table\ 
\ref{tab2} is because of a rather rough numerical method of 
extracting $\Delta E^{(+)}_{0+}$ and $\Delta p^{(+)}_1$.  
Nevertheless, we believe that such a method properly reproduces 
the tendency of variation of $E^{(+)}_{0+}$ and $p^{(+)}_1$ due 
to $\pi$-$N$ rescattering. The dependence of  
$\Delta E^{(+)}_{0+}$ and $\Delta p^{(+)}_1$ on the choice of nucleon 
energies as seen in Table\ \ref{tab2} is expected  since the
Cases 1 to 3 effectively lead to different pion propagators in Eq.\ 
(\ref{np-pin-resc}).

Averaging over all values in Table\ \ref{tab2} we obtain $\Delta 
E^{(+)}_{0+}\simeq -1.1$.  One can see that $\Delta E^{(+)}_{0+}$ 
is negative in accordance with the ChPT result from Ref.\ 
\cite{beane} being, however, nearly two times smaller in the 
absolute value.  The $p^{(+)}_1$ amplitude acquires a positive 
contribution an averaged value for which is $4.1 \pm 0.6$.  This 
result being put together with the free-nucleon value of about 9 
gives $p_1=13.1 \pm 0.6$ and reasonably reproduces an experimental 
result of Ref.\ \cite{bergstrom}, $p^{exp}_1=12.88\pm 0.28$. In other 
words, our estimates confirm an assumption from that paper that a 
mechanism responsible for renormalizing the $p_1$ amplitude is the 
two-body process corresponding to $\pi$-$N$ rescattering.

\begin{table}[hbt]
\caption 
{
Effective modifications of the threshold $E^{(+)}_{0+}$ and 
$p^{(+)}_1$ amplitudes due to $\pi$-$N$ rescattering. 
Units are conventional.
}
\begin{center}
\begin{tabular}{l|cccccc}
\multicolumn{1}{l|}{} &
   \multicolumn{2}{c} {Case 1} 
 & \multicolumn{2}{c} {Case 2} 
 & \multicolumn{2}{c} {Case 3} \\
     &$\Delta E^{(+)}_{0+}$ &$\Delta p^{(+)}_1$ 
     &$\Delta E^{(+)}_{0+}$ &$\Delta p^{(+)}_1$ 
     &$\Delta E^{(+)}_{0+}$ &$\Delta p^{(+)}_1$ \\
\hline
ChPT & $-1.2$    & +3.7   & $-1.1$   & +4.2   & $-1.2$   & +4.7   \\
DR   & $-1.0$    & +3.5   & $-1.1$   & +3.8   & $-1.1$   & +4.5   \\
\end{tabular}
\end{center}
\label{tab2}
\end{table}

To make it much easier using our results in practice,
we have parametrized the differential cross sections at four 
photon energies 145, 150, 155, and 160 MeV with the help of the 
following formula
\beqn
\frac {d\sigma}{d\Omega_\pi}= A_1+A_2\cos {\Theta_\pi},
\label{fit}
\eeqn
with parameters given in Table\ \ref{tab3}. The values for 
$A_1$ and $A_2$ have been obtained by a fit to the average of the   
minimum and the maximum cross sections (see above) 
and the errors in $A_1$ account for variations of the cross sections 
both due to various choices of sets for multipoles and nucleon 
energies in the right loop of diagram \ref{fig1}{\it f}.  
The corresponding cross sections at arbitrary energies up to 160 MeV can 
be  obtained by making use of a suitable interpolation procedure.   
We think that such a simple parametrization provides reasonable 
accuracy to be used in estimates of inelastic channel 
contributions to coherent $\pi^0$ production.

\begin{table}[hbt]
\caption 
{
Coefficients $A_1$ and $A_2$ in Eq. (\ref{fit}) at four selected 
photon energies. Units are  $\mu$b/sr.
}
\begin{center}
\begin{tabular}{l|cccc}
      & 145 MeV & 150 MeV  & 155 MeV  & 160 MeV \\
\hline
$A_1$ & 0.0043 $\pm$ 0.0014  & 0.023 $\pm$ 0.005
      & 0.057 $\pm$ 0.007  & 0.095 $\pm$ 0.012 \\ 
$A_2$ &$-0.0007$  & $-0.007$  & $-0.026$  & $-0.041$ \\ 
\end{tabular} 
\end{center}
\label{tab3}
\end{table}

\begin{figure}[thb]
\epsfxsize=0.5\textwidth
\centerline{\epsfbox[30 330 480 810]{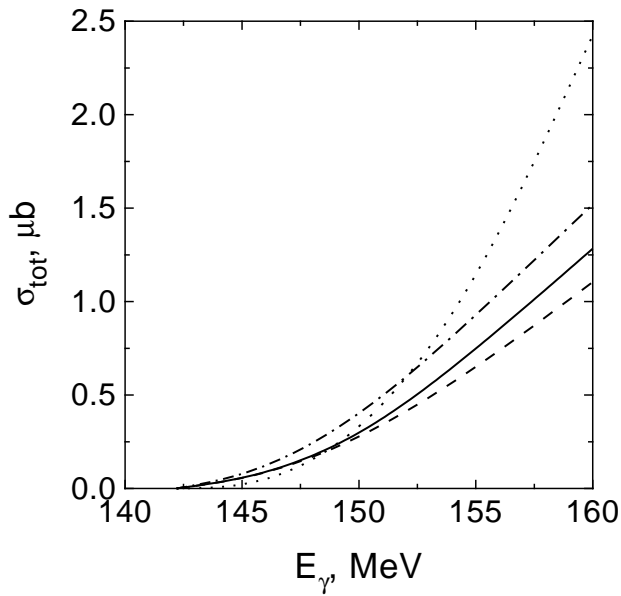}}
\caption
{
Total cross section of the reaction  $^2H(\gamma,\pi^0)np$ in the 
threshold region as a function of photon lab energy.
Meaning of curves as in Fig. \ref{fig2}.
}
\label{fig4}
\end{figure}

Performing in Eq.\ (\ref{dcs}) an integration over the pion solid 
angle $\Omega_\pi$ we can calculate the total cross section 
$\sigma_{\rm tot}$ of the reaction (\ref{react}) (note that 
$\sigma_{\rm tot}=4\pi A_1$ where $\sigma_{\rm tot}$ is in units of 
$\mu$b).  In Fig.\ \ref{fig4} we  show contributions of different 
diagrams to $\sigma_{\rm tot}$ with the ChPT set for the  multipoles 
in Eq.\ (\ref{benmer}). The two pole diagrams produce a total cross 
section which is rapidly increasing with photon 
energy.  After inclusion of the $n$-$p$ final state interaction the 
energy distribution of $\sigma_{\rm tot}$ becomes to be more flat. 
Just above threshold this interaction in the $^1S_0$-wave leads to 
an increase of $\sigma_{\rm tot}$ but at $E_\gamma \geq 155$ MeV the 
cross section diminishes by a factor of about 2.  Diagrams 
\ref{fig1}{\it d} and {\it e} with $\pi$-$N$ rescattering give 
noticeable positive contribution to the total cross section. Of all 
the diagrams \ref{fig1}{\it f} is the least important one, decreasing 
$\sigma_{\rm tot}$ by about 10\%.

\begin{figure}[thb]
\centerline{
\leavevmode\epsfxsize=0.50\textwidth \epsfbox{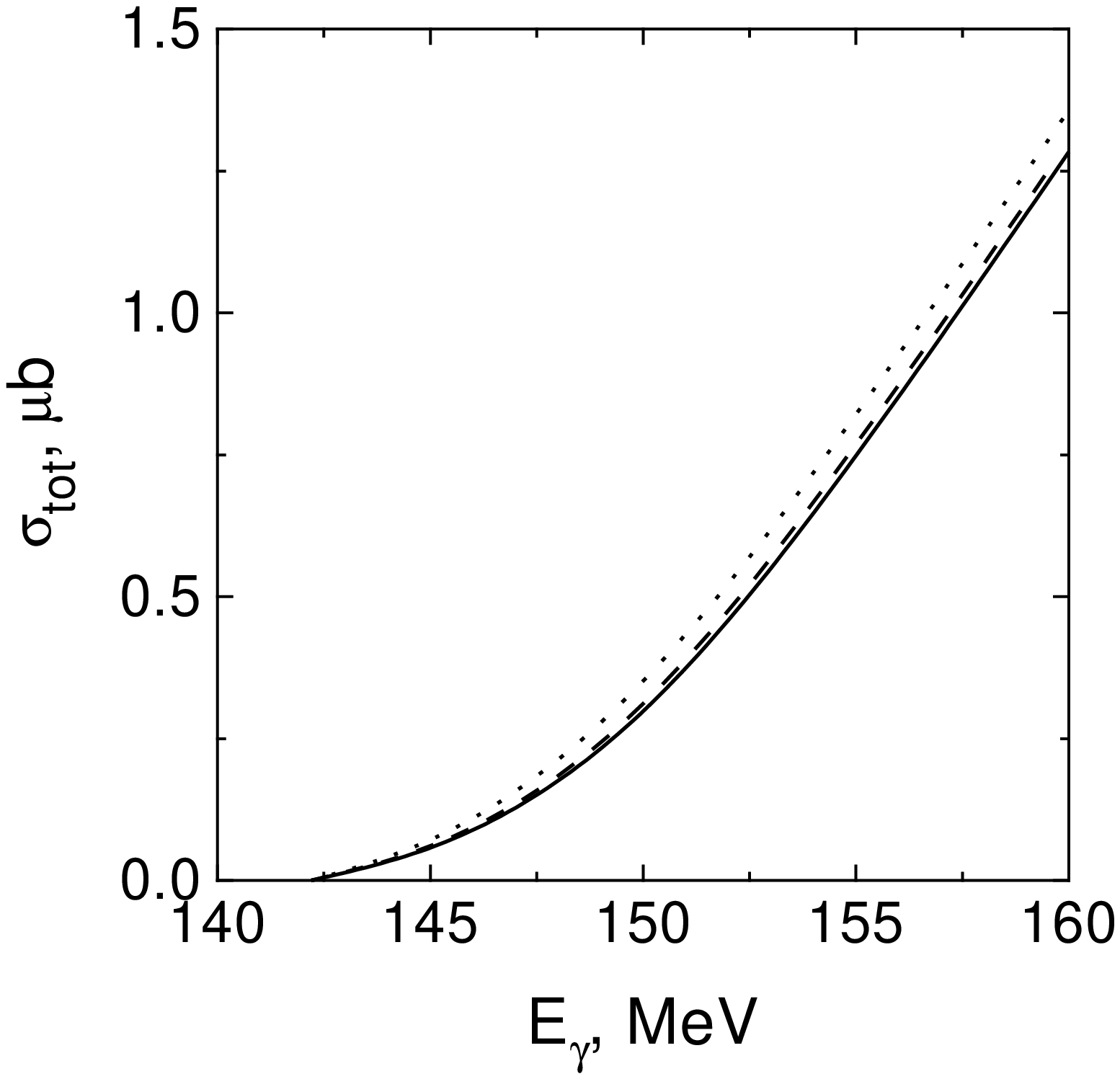}
\leavevmode\epsfxsize=0.50\textwidth \epsfbox{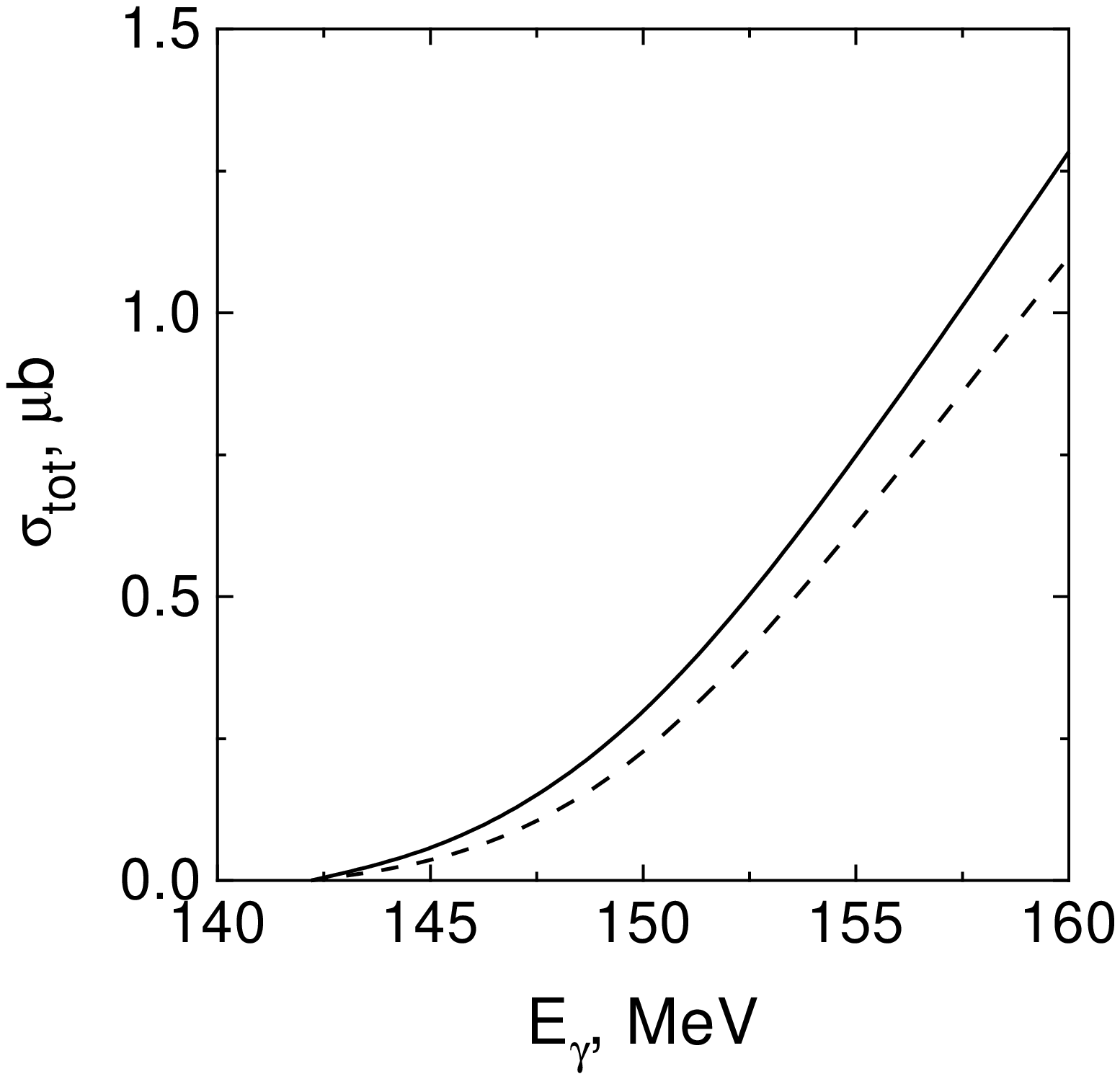} }
\caption
{
Left panel: shown is the sensitivity of the total cross section to  
different prescriptions for nucleon energies in the right loop of 
diagram \ref{fig1}{\it f}.
Meaning of curves as in Fig. \ref{fig3_1}.
Right panel:
sensitivity of the total cross section to sets 
of the multipoles in Eq. (\ref{benmer}). 
Meaning of curves as in Fig. \ref{fig3}.
}
\label{fig5}
\end{figure}

The total cross sections for various choices of nucleon energies in 
the right loop of diagram \ref{fig1}{\it f} are presented in the left 
panel of Fig.\ \ref{fig5}. It is seen that they are  practically 
independent of a prescription for the energy. This reflects 
the smallness of the diagram \ref{fig1}{\it f} 
contribution to the total cross section.  Our predictions for 
$\sigma_{\rm tot}$ with the ChPT and DR sets of the multipoles are 
also shown in Fig.\ \ref{fig5}. There is a deviation of about 15\% in 
the full energy region. 

\begin{figure}[hbt]
\centerline{
\leavevmode\epsfxsize=0.50\textwidth \epsfbox{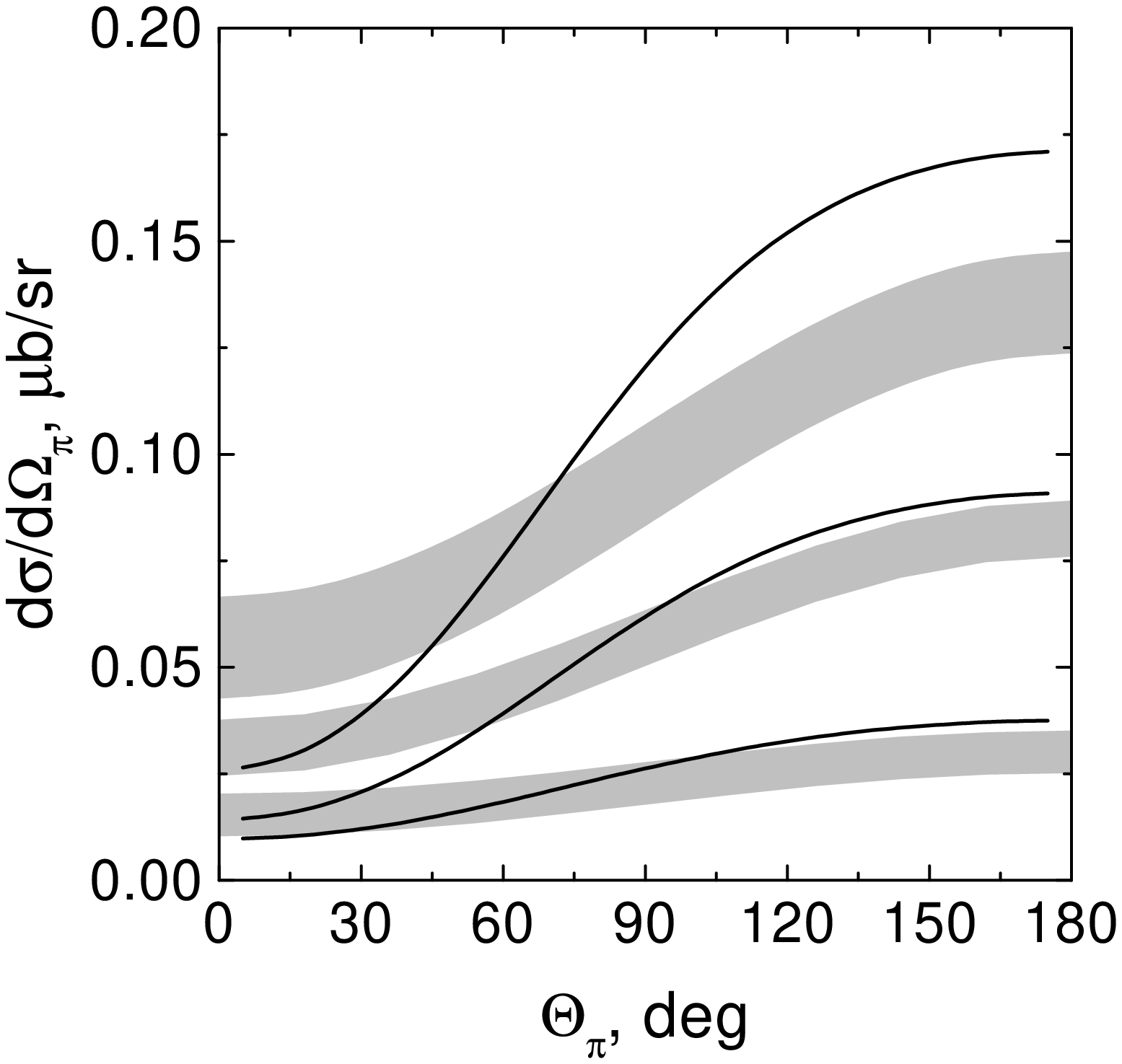}
\leavevmode\epsfxsize=0.50\textwidth \epsfbox{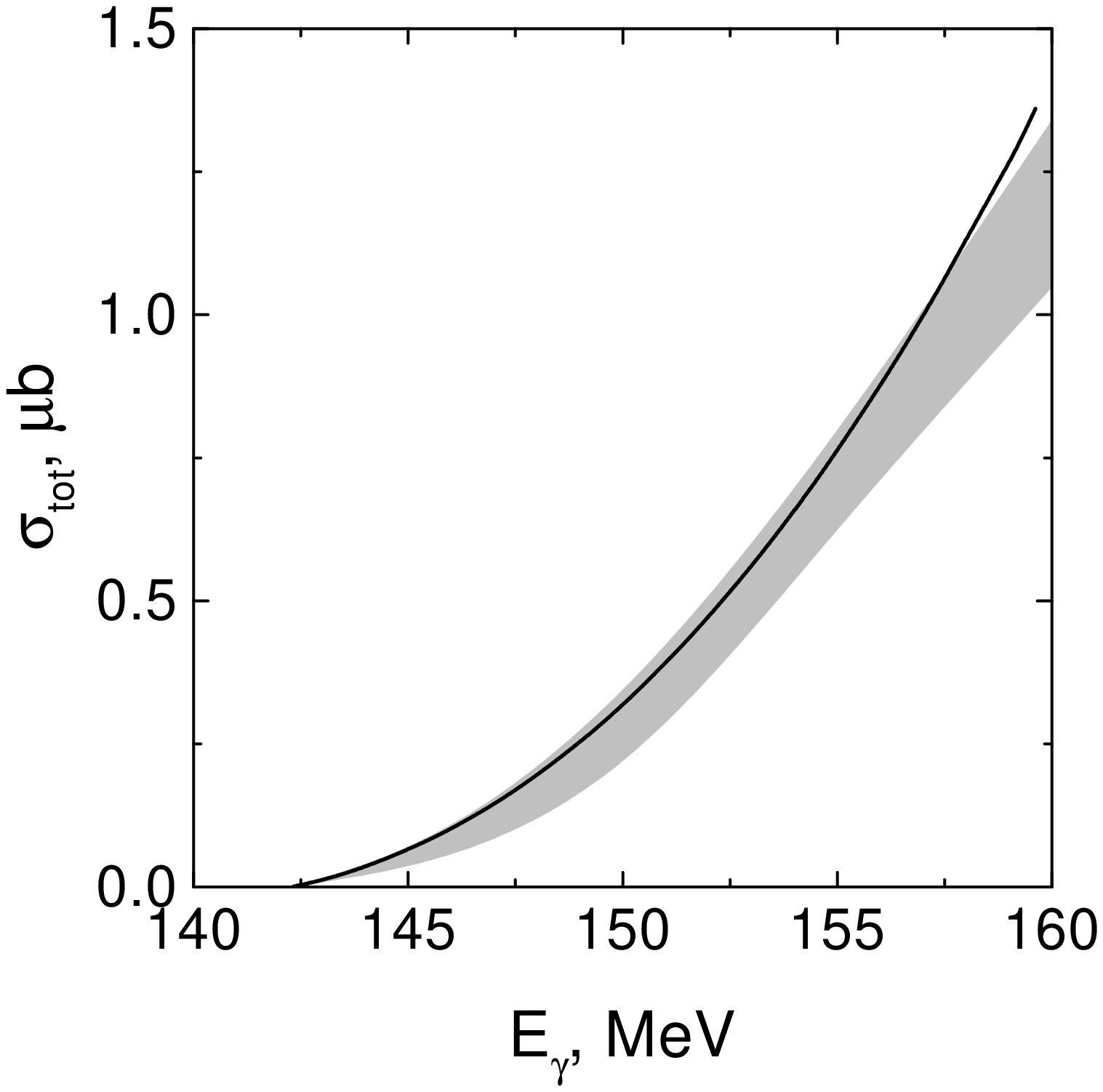} }
\caption{
Left panel: shown are errors bars (filled areas) in the 
differential cross sections stemming from various 
prescriptions for nucleon energies and different sets of 
multipoles at 150, 155, and 160 MeV (from bottom to 
top) calculated with Eq. (\ref{fit}) and coefficients $A_1$ and 
$A_2$ from Table \ref{tab3}. Curves are results of Ref.\
{\protect \cite{bergstrom}}.  Right panel: the same as for the left
panel but for the total cross section.
}
\label{errors}
\end{figure}

In Fig.\ \ref{errors} we compare our results with the  ones 
of the model \cite{bergstrom} for the differential and total cross 
sections.  One can see that the predictions of the two calculations 
are quite consistent. There is some disagreement  at $E_\gamma \ge 
155$ MeV and forward angles for the differential cross section.  It 
should be emphasized, however, that this disagreement cannot make any 
noticeable changes in the cross sections of coherent $\pi^0$ 
photoproduction measured in Ref.\ \cite{bergstrom} because of 
smallness of the inelastic channel cross sections in comparison with 
the coherent channel ones.  
At the same time, at backward angles the 
inelastic channel is growing in importance but there the deviation is 
not very significant.  One can see  that the maximum deviation between 
the present calculation and the one of  Ref.\ \cite{bergstrom} does not 
exceed 20\% at 160 MeV and backward angles, being even smaller when the 
energy is decreasing. 
(Note here that the authors of that paper ascribed to their calculated 
cross sections of the inelastic channel uncertainties of about $\pm 25\%$ 
which were taken into account in their data analysis).  
In the case of the total cross section we observe good agreement with 
Ref.\ \cite{bergstrom}. 
An analysis  performed by Bergstrom \cite{bergstrom_pc} has shown 
that  the $p_1$ amplitude is insensitive to the variation in the 
inelastic cross section from the present work and that from Ref.\ 
\cite{bergstrom}.  Likewise, the extrapolation of the $E_d$ amplitude 
to threshold changes very little, remaining within the error in Eq. 
(12) of Ref.\ \cite{bergstrom}.

In summary, we have performed an investigation of the  reaction 
$^2H(\gamma,\pi^0)np$ in the threshold region and found a good 
agreement with the only previous calculation \cite{bergstrom} for 
the total cross section. A reasonable agreement also exists for  the 
differential cross sections. Our predictions coincide within of 
20\%. If  one takes into account that the authors of that work 
ascribed to their results an uncertainty of about $\pm 25\%$ 
we conclude that the present calculation cannot lead to any 
changes in the total and differential cross section of coherent 
$\pi^0$ photoproduction on the deuteron measured in Ref.\ 
\cite{bergstrom}.  This means that the deviation of about 20\% from 
theoretical predictions found in that paper for the $E_d$ threshold 
amplitude still remains to be resolved.  Our calculations have shown 
that the difference of about $4\cdot 10^{-3}/\mu^3_{\pi^+}$ between 
the free-nucleon value for the $p_1$ threshold amplitude and 
the experimental value can be attributed to two-body effects due to 
$\pi$-$N$ rescattering.

{\acknowledgments 
We would like to thank J.C. Bergstrom for supplying us with 
results of his numerical calculations and for fruitful discussions.  
We are pleased to acknowledge helpful comments by  A.I. L'vov. One of 
the authors (M.L.) is indebted to M.V.  Galynsky for the use of his 
computer.  This work was supported by Advance Research Foundation of 
Belarus and Deutsche Forschungsgemeinschaft under contract 
436 RUS 113/510.}

\end{document}